%% file: main.tex
\documentclass[letterpaper,twocolumn,10pt]{article}
\usepackage{usenix}

\usepackage{tikz}
\usepackage{amsmath}
\usepackage{graphicx}
\usepackage{soul}
\usepackage{xcolor}
\usepackage{enumitem}
\usepackage{array}
\graphicspath{{figures/}}

\newenvironment{tightitemize}
  {\begin{itemize}[noitemsep, topsep=2pt, parsep=0pt, partopsep=0pt]}
  {\end{itemize}}

\newenvironment{tightenumerate}
  {\begin{enumerate}[noitemsep, topsep=2pt, parsep=0pt, partopsep=0pt]}
    {\end{enumerate}}

\newif\ifanonymous
\anonymousfalse

\input{figures/soc_ticket_style}

\begin{document}
%-------------------------------------------------------------------------------

%don't want date printed
\date{}

\title{\Large \bf It is Not Yet Another Tool: Creating and Deploying an Agentic AI Companion in a Security
  Operations Center}

%for single author (just remove % characters)
% \author{
% {\rm Your N.\ Here}\\
% Your Institution
% \and
% {\rm Second Name}\\
% Second Institution
% % copy the following lines to add more authors
% % \and
% % {\rm Name}\\
% %Name Institution
% } % end author

% \author{} % anonymized for submission

% Fixed-width author grid so the two rows line up in the same three columns
\newcommand{\authblock}[3]{%
  \begin{minipage}[t]{0.32\textwidth}\centering
  {\rm #1}\\ #2\\ #3
  \end{minipage}}

\author{
\authblock{Kritan Banstola\thanks{Banstola and Faisal share first authorship}}{University of South Florida}{kbanstola@usf.edu}%
\authblock{Faayed Al Faisal\footnotemark[1]}{University of South Florida}{faayedal@usf.edu}%
\authblock{Duy Dao}{Cyber Florida}{duydao@usf.edu}\\ \noalign{\vskip 1.2ex}
\authblock{Ryan Irving}{Cyber Florida}{irvingr@usf.edu}%
\authblock{Daniel Lende}{University of South Florida}{dlende@usf.edu}%
\authblock{Xinming Ou}{University of South Florida}{xou@usf.edu}
} % end author

% keep the \thanks marker black; hyperref would otherwise color it as a link
\hypersetup{linkcolor=black}
\maketitle
\hypersetup{linkcolor={green!80!black}}

%-------------------------------------------------------------------------------

\input{abstract}

\input{introduction}
\input{related_work}
\input{methodology}
\input{design}
\input{data-analysis}
\input{discussion}
\input{conclusion}
\input{acknowledgement}
% --- end template filler ---

%-------------------------------------------------------------------------------
% optional clearing of the page
\cleardoublepage
\appendix
\input{ethical_considerations}
% optional clearing of the page
% \cleardoublepage
\newpage

\input{open_science}
% optional clearing of the page
\cleardoublepage

\bibliographystyle{unsrturl}
\bibliography{soc_ai,anth}

\input{appendix}

%%%%%%%%%%%%%%%%%%%%%%%%%%%%%%%%%%%%%%%%%%%%%%%%%%%%%%%%%%%%%%%%%%%%%%%%%%%%%%%%
\end{document}

%% file: figures/soc_ticket_style.tex
\usepackage[most]{tcolorbox}
\usepackage{xcolor}

\definecolor{ticketbackground}{HTML}{FAF8F5}
\definecolor{ticketborder}{HTML}{E5D8C5}

\newcommand{\anon}[1]{{\normalfont\itshape$\langle$#1$\rangle$}}

\newtcolorbox{ticketpanel}[1][]{
  enhanced,
  colback=ticketbackground, colframe=ticketborder,
  boxrule=0.8pt, arc=3pt,
  left=5pt, right=5pt, top=5pt, bottom=5pt, boxsep=0pt,
  #1
}

%% file: abstract.tex
\begin{abstract}
%-------------------------------------------------------------------------------
  Security Operations Centers (SOCs) process large amounts of tickets, most of which are
  low-interest events not worthy of further investigation. The repetitive nature
  of this task and similarity of the vast amounts of tickets make it a prime candidate
  for generative AI-based automation. We created and deployed an agentic AI companion
  utilizing large language models through fieldwork within a SOC for over one year.
  The design of the SOC AI companion was driven by researchers' participation and interactions
  within the SOC's daily work. SOC analysts were invited to use it during
  the last four months of the fieldwork. We analyzed the
  analysts' usage of the companion and found that in more than 90\% of the cases
  the companion's outputs were reused by analysts in the ticket's closing report.
  Our results showed that when designed ``in the trenches'' with the intended
  users, a SOC AI companion can go beyond being yet another tool, but rather
  a system that co-evolves with its human users as it traverses through the various types
  of workloads. Analysts naturally
  started to shape the AI companion's behaviors to fit their particular needs.
  Our data show that the more human analysts shape the AI companion's
  behaviors, the more they become comfortable trusting the output from the AI system, resulting
  in improved productivity.
\end{abstract}

%%% Local Variables:
%%% mode: latex
%%% TeX-master: "main"
%%% End:

%% file: introduction.tex
%-------------------------------------------------------------------------------
\section{Introduction}
\label{sec:intro}
%-------------------------------------------------------------------------------
A Security Operations Center (SOC) is responsible for monitoring an organization's
networks and systems for evidence of compromise.
Most of that responsibility is discharged through alert triage.
Detection systems produce alerts continuously, and analysts work through them
deciding whether an alert indicates a genuine intrusion
or a benign event that happens to satisfy a detection rule.
Most of the cases are benign~\cite{Alahmadi2022Usenix};
however an analyst must still work through the alert to make that determination
which involves assembling context by navigating through various tools and data sources.
This process is monotonous and laborious, and it contributes to fatigue
and attrition among analysts~\cite{Sundaramurthy2015Usenix,Shropshire2012CiteSeer}.
Generative AI, in particular large language models (LLMs) is potentially suited
to provide automation that can alleviate this burden due to its capability to
reason through varied situations presented by data. In fact analysts are already
actively using LLMs as a chatbot assistant when processing tickets~\cite{Singh2025IEEE,
  Nath2026Usenix,Wong2026SOUPS}.
A natural next step is to create an agentic AI companion
so it can take actions on analysts' behalf such as querying the various SOC tools
and data sources to establish context. While there have been multiple SOC tool
vendors~\cite{ReliaQuest2018ReliaQuest, Rapid72025Rapid7, CrowdStrike2026AIDR}
who claimed to have incorporated agentic AI in their solutions, there is little
documentation beyond marketing materials on how these AI SOC assistants operate. Those commercial
AI-enabled solutions are marketed as yet another tool/service a SOC can purchase.
However there are a number of significant differences between an agentic AI
system and traditional SOC tools that would create significant hurdles for
agentic AI solution's adoption by SOCs.

First and foremost, analysts must be able to trust an agentic AI companion's
reasoning. It is well known that LLMs' reasoning could be wrong.
Ultimately a human analyst must close a SOC ticket and bear the responsibility
on the decisions made, regardless of whether AI provided incorrect information
or not. % This is very different from using a non-AI tool, which would have
% well defined input and output semantics and will only be wrong when there is a
% bug in the tool, in which case the analyst would not be held responsible.
Thus it is unlikely a SOC analyst will readily accept an AI system's
output without verification. If the effort spent on verification exceeds
the effort of solving the ticket manually it defeats the purpose of automation.

Second, if an agentic AI companion is programmed like yet another tool for the SOC, like
all tools it is bound to fall short to fit some aspects of a specific SOC's
situation. This could be peculiar data sources the SOC relies upon, the
specific business and risk environment where the SOC is monitoring, subtle
workflow differences dependent on organization structure, and so on. Traditional
SOC solutions' one-size-fits-all approach is notorious for not providing the needed
adaptability and that is one main reason behind the difficulty of automating
the mundane SOC tasks that create burnout for analysts~\cite{Shropshire2012CiteSeer}. 
% Given generative AI's capability of understanding human language,
An agentic AI
assistant for SOC should allow users to shape its behaviors
so it can address the specific situation a SOC faces.
Such an AI assistant would be a companion to the SOC analysts that can co-evolve
with the analysts and the work environment as they change, as opposed
to yet another rigid, hard-to-modify tool.

In this paper we present a 14-month long fieldwork study of a SOC located
in an institution of higher education.
%Our anthropological study followed that of Sundaramurthy~\cite{Sundaramurthy2015Usenix}.
Two PhD student researchers
became the SOC's analysts and used the research method of participant
observation~\cite{Bernard2011AltaMira,Agar1996Emerald}
to analyze the work conducted by the SOC's analysts. Throughout the research we
identified opportunities where agentic AI solutions can help automate the mundane
part of the workflow, and created an AI companion to achieve that based on the
researchers' first-hand understanding of the work themselves.
The SOC AI Companion is a ReAct-style LLM agent~\cite{Yao2023ICLR} that can independently
run an investigation on a SOC alert.
It runs on a locally deployed model, %so alert data never leaves the SOC's network.
reads in each ticket and follows the same workflow the analysts are asked to follow: 
queries the SOC's log sources, enriches indicators against internal and external services,
and drafts a structured report for the analyst to review.
All evidence is presented in the report in the same way as would by a manual analysis.
We then invited the other SOC analysts to use the AI companion for the last
four months of the fieldwork. 
Analysts interact with the companion through an interface that allows for
asking questions, providing additional context, and changing system prompts. 
Analysts used it on real alerts as they came in and the use was voluntary.
The AI companion records all transcripts and actions between users and the AI model.
We also retain data in the ticketing system that show how the analysts processed
the ticket in the end -- did they use the AI companion's results or not?
We then carried out a systematic analysis of the collected data along with
the researchers' field notes to examine the following
key aspects of creating and deploying agentic SOC AI companions in SOCs.

% \begin{itemize}
%     \item \textbf{RQ1.} Can an Agentic AI system do end-to-end investigation for a SOC alert?
%     \item \textbf{RQ2.} What does it take for an AI companion to be adopted into the daily work of a SOC?
%     \item \textbf{RQ3.} What improvements can an AI companion bring to an analyst workflow in a SOC?
% \end{itemize}

\begin{tightitemize}
\item \textbf{RQ1.} Would SOC analysts % trust the AI companion; in other words, would they
  use the AI
  companion's analysis and recommendation to close tickets? 
\item \textbf{RQ2.} Do analysts attempt to modify the SOC AI companion's behavior to fit better
  into their work?
\item \textbf{RQ3.} What improvements, or lack thereof, the SOC AI companion brought to the SOC
  analysts' work?
\end{tightitemize}

Our main findings and contributions are:

\begin{tightenumerate}

\item We found that when the SOC companion's output is presented in an easy-to-verify way,
  SOC analysts overwhelmingly used its results in ticket closing reports.
  It is not sufficient to have accurate results; the biggest impact on adoption appears to be
  verifiability. % Not only there are both explicit and implicit trust signals that impact whether
  % a SOC analyst would use our SOC companion's investigation results. The explicit signal
  % refers to the accuracy of the SOC companion's analysis; the implicit signal refers to
  % the forms of how the results and evidence are presented.
    Our fieldworkers were able to observe the importance of this implicit signal and adapted
    the AI companion's output, resulting in a much higher adoption rate.
  
\item We found that analysts naturally started attempts to change the AI companion's behaviors
  to better fit into their needs and habits.
  With help from our fieldworkers,
  analysts became well versed in using personalized system prompts that 
  encoded their own investigation routines and preferred report layouts. 
  The ability to adapt the AI companion also
  resulted in higher adoption rate.

\item We measured the time it took for analysts to close SOC tickets using the AI SOC companion,
  and compared that with self-reported manual analysis time. The SOC
  companion saved an estimated $30-50\%$ of ticket-processing time for the most prolific Companion users.
  Analysts also reported that the time spent on closing a ticket using our AI companion is perceived
  to be more productive than if the ticket is handled manually. 
  
\end{tightenumerate}

Our findings indicate that to make a productive AI assistant for SOCs, its design can follow
a different process than a traditional SOC tool. Instead of embedding AI inside yet another tool
and asking analysts to orient their workflow around that system, one can design an AI companion
{\it based on} the existing workflow in the SOC, so that it can immediately offload burdens from the
analysts. %instead of creating yet another new set of signals they have to work on.
Verifiability is the key to adoption; an AI companion that produces its evidence in the same
format as the manual process would is much more likely to pass muster and be adopted by analysts.

% To that end
% it is helpful to expose key elements of an AI system to users (analysts) through a proper
% abstraction, so the analysts with simple instruction can shape the AI companion's behaviors based
% on the unique environment of the SOC they work in.

% \bigskip
% \noindent\textcolor{red}{\textbf{TO DO:} summary of study design ,summary of the data analysis and findings, followed by the contributions of the paper.}
% \bigskip

% The remainder of this paper is organized as follows.
% Section~\ref{sec:related_work} reviews related work. %  on how analysts work in security operations
%     % and on the use of LLMs for security tasks.
% Section~\ref{sec:methodology} describes research methodology.  % our fieldwork within an operational SOC,
%     % through which we study the design, deployment and continued use
%     % of the SOC AI Companion.
% Section~\ref{sec:design} discusses the design and deployment of the SOC AI Companion.
% Section~\ref{sec:analysis} presents the data analysis and findings.
% Section~\ref{sec:discussion} offers some discussions from the findings.
% % some the implications of these results
% %     for the design of AI assistance in security operations,
% %     along with the limitations of our study.
% Section~\ref{sec:limitations} discusses the limitations of the research
% and how we mitigate them.
% Finally, Section~\ref{sec:conclusion} concludes.
% % the paper
% % and outlines future directions for research in this area.

%%% Local Variables:
%%% mode: latex
%%% TeX-master: "main"
%%% End:

%% file: related_work.tex
\section{Related Work}
\label{sec:related_work}

A number of recent works examine LLM's use directly within SOC and
incident-response contexts. Singh et al.~\cite{Singh2025IEEE} analyze
3,090 queries from 45
SOC analysts over 10 months, finding that analysts primarily use LLMs
as on-demand aids for sensemaking, interpreting low-level telemetry,
and refining technical communication, while preserving analyst
decision authority. Kramer et al.~\cite{Kramer2025Usenix} study LLMs for
incident-response summarization using 18 analysts and 50 real-world
incidents, showing that fully autonomous summaries often omit critical
details or introduce factual inaccuracies, while collaborative
LLM-assisted summaries can reduce analyst effort and improve
readability and consistency. Other studies
using Reddit discussions~\cite{Nath2026Usenix}
and interviews with 28 cybersecurity professionals across 26
organizations~\cite{Wong2026SOUPS} find substantial LLM use
alongside concerns about reliability, verification, security or data
leakage, and organizational guidance.
Our work complements these studies by going through the full circle
of designing, implementing, deploying, and observing the usage of
an LLM-based agentic AI assistant within an operational SOC while
researchers working inside the SOC as analysts.
A unique strength of our approach is the research method
of participant observation, which allows us to observe the SOC's
work in zero proximity and obtain first-hand data about how analysts
used and interacted with an AI companion.

Recent work has explored the use of AI and LLMs across a wide range of
cybersecurity tasks, including penetration testing, phishing
detection, vulnerability analysis, alert classification, and
SOC triage~\cite{Deng2024Usenix,
  Koide2024IEEE,Kasri2025MDPI,
  Khayat2025IEEE, Freitas2025ACM,Turcotte2025Arxiv, Banstola2026WOSOC}.
What differs this work from those prior efforts is that we evaluate our approach
inside a working SOC by observing how analysts used and interacted with an
agentic AI assistant.

Our work has been informed by a large body of earlier research
in security operations.
Recent studies document false-positive and alert-validation burden,
SOC analyst burnout, broader human, organizational, tooling
challenges, and importance of analysis structure~\cite{Jones2023Usenix,
  Alahmadi2022Usenix, Sundaramurthy2015Usenix,
  Werlinger2009Emerald,Werlinger2008ACMb,Kokulu2019CCS,Kersten2023SOUPS}.
Earlier field and human-centered studies
show that security work
depends on tacit knowledge, context, and organizational
practice~\cite{Jaferian2008ACM,Botta2007ACM}.
They also emphasize that security tools should be designed and evaluated in
relation to analysts' existing activities and constraints, with
practitioners involved in how those tools are developed and
refined.
% These research shows that SOC challenges are shaped not only by
% the capabilities of detection
% systems, but also by organizational procedures, analyst expertise, and
% the practical knowledge required to conduct
% investigations.
% Related work has also examined structured support for Tier-1
% investigation workflows~\cite{Kersten2023SOUPS}.
 %~\cite{Werlinger2010Emerald,Botta2011Springer,Sundaramurthy2014IEEE,Sundaramurthy2016Usenix}.
 % ~\cite{Suchman1987Cambridge,Polanyi1966Doubleday,Lave1991Cambridge,Turner1994Chicago}.
%~\cite{Jaferian2008ACM,Botta2007ACM,Czyzewski1990PDC,Squires2002Greenwood,Vines2013ACM}
These prior works motivate our work in evaluating AI assistance as part of everyday SOC
practice rather than only as an isolated technical capability.

% Such architectures are well
% suited to SOC investigations, which often require analysts to move
% between hypotheses, logs, threat-intelligence sources, and reporting
% tasks.

% Much of these works, however, evaluates LLMs as task performers or
% technical components. Automated alert-classification and
% guided-response systems show that AI can prioritize alerts or
% recommend actions~\cite{Turcotte2025Arxiv}, while
% systems such as PentestGPT demonstrate how LLMs can assist with
% structured, multi-step security tasks~\cite{Deng2024Usenix}.

%%% Local Variables:
%%% mode: latex
%%% TeX-master: "main"
%%% End:

%% file: methodology.tex
%-------------------------------------------------------------------------------
\section{Methodology}
\label{sec:methodology}
%-------------------------------------------------------------------------------

% Our goal was to study how an agentic AI assistant could be designed
% for, deployed within, and incorporated into everyday SOC triage
% work.
% rather
% than a one-time laboratory evaluation. The study proceeded in three
% phases:
We conducted a longitudinal, embedded field study in a SOC.
The study consisted of three concurrent efforts:
(1) the research method of participant
observation~\cite{Bernard2011AltaMira,Agar1996Emerald}
within the SOC to learn, observe, and understand the SOC's work,
(2) iterative design and development of a SOC AI
companion based on the observed workflow and analyst feedback,
and (3)
deployment and analysis of analysts' use of the AI companion during live
ticket work.

\paragraph{Research site}

The study was carried out over approximately fourteen months in the
SOC of a large university. The SOC monitors institutional systems and
networks, with analysts primarily performing tier~1 triage: reviewing
alerts, gathering evidence from security tools and logs, determining
whether activity is benign or requires escalation, and documenting the
result in the ticket. Triage required analysts to combine formal
procedures with local knowledge of the organization's systems, users,
and recurring alert patterns.
Most of the analysts were recruited from the university's student body
and it was a highly competitive job to land. Each student analyst spent
one to two years in the SOC, working 10-20 hours a week on operational tasks.
% and gaining
% real-world security
% operation knowledge and skills.
Thus the SOC not only carried its operational duties but also provided
an experiential learning environment for the student analysts.
At the time of our fieldwork, there were between 10 to 20 student analysts
in the SOC at any point, two full-time analysts, and two managers overseeing the
day-to-day operations.

% The study was carried out in a SOC inside a large
% university and lasted approximately fourteen months.
% The SOC monitors institutional systems and networks and
% processes alerts through a ticketing platform. Analysts primarily
% perform Tier 1 triage: reviewing alerts, gathering relevant context
% from security tools and logs, determining whether the activity is
% benign or requires escalation, and documenting the result in the
% ticket.
% This setting was suitable for studying AI assistance because much of
% the work involved repetitive evidence gathering across disconnected
% tools, while still requiring situated analyst judgment. For example,
% analysts often needed to determine which host or user was involved in
% an alert, whether the behavior matched known benign activity in the
% local environment, whether related events appeared in logs, and
% whether the available evidence was sufficient to close or escalate the
% ticket. These tasks were routine, but not trivial. They required
% analysts to combine formal playbooks with local knowledge of the
% organization, its systems, and its recurring alert patterns.

\noindent{\bf Research ethics and IRB} Prior to carrying out the fieldwork,
this research was reviewed and approved by the institutional review
board (IRB) of the university. Verbal informed consent was conducted to
every employee of the SOC and participation was strictly voluntary. Detailed
ethical considerations are discussed in the standalone section towards
the end of the paper.

\subsection{Participant observation}
\label{sec:participant_observation}

Participant observation is an ethnographic method in which researchers
learn a practice by participating in it and observing how work is
actually performed in context~\cite{Agar1996Emerald,Bernard2011AltaMira}.
Past research has adopted this method in studying SOCs~\cite{Sundaramurthy2015Usenix}
and we follow a similar approach in our study.
Two PhD student researchers joined the SOC as tier 1 analysts. They
were trained on the SOC’s tools and procedures and worked regular shifts
alongside other analysts.
This embedded role allowed the researchers to learn the work by
doing it rather than relying solely on documentation, interviews,
or post-hoc explanations.
Moreover, by working tickets as regular analysts, our researchers
became seen as part of the team. This is the key factor in 
gaining and maintaining trust which is critical for studying SOCs~\cite{Sundaramurthy2015Usenix}. 

% Participant observation is an ethnographic method in which researchers
% learn a practice by participating in it and observing how work is
% actually performed in context~\cite{Agar1996Emerald,Bernard2011AltaMira}.
% Past research has adopted this method in studying SOCs~\cite{Sundaramurthy2015Usenix}
%  and we follow a similar approach in our study.
% Two PhD student researchers joined the SOC as tier 1 analysts. They
% were trained on the SOC's tools and procedures and worked regular shifts
% alongside other analysts.
% % and handled the same types of alerts that appeared in the SOC
% % queue.
% This embedded role allowed the researchers to learn the work by
% doing it rather than relying solely on documentation, interviews, or
% post-hoc explanations.
% Gaining and maintaining trust is critical
%   for participant observation and such trust is hard to earn in SOCs~\cite{Sundaramurthy2015Usenix}.
% Our researchers achieved this
% by working tickets as regular analysts.

Throughout the fieldwork, the researchers maintained field notes
documenting routine triage practices, sources of delay, tool-switching
behavior, uncertainty during investigations, informal analyst
feedback, and potential opportunities for automation. The notes
included both direct observations of SOC work and reflective memos
about emerging patterns. Via collaborative analysis, these observations
and reflections became the basis for the design decisions that followed.
The data collected, while limited to one site, go deep into the daily
practices of a practicing SOC and what works in real-time for both
analysts and workflows. It thus offers an important complement to
cross-site data which are more comparative but also more
ethnographically thin because there is not as much verification
against actual work practices such as our research approach provides.

% Throughout the fieldwork, the researchers maintained field notes
% documenting routine triage practices, sources of delay, tool-switching
% behavior, uncertainty during investigations, informal analyst
% feedback, and potential opportunities for automation. The notes
% included both direct observations of SOC work and reflective memos
% about emerging patterns.
% % Through participant observation and field notes, we found that formal
% % playbooks captured only part of analysts' work. Analysts routinely relied on
% % local knowledge, informal practices, and experience when deciding how to
% % interpret alerts and when sufficient evidence had been gathered.
% These observations and reflections became the basis for the design decisions
% that followed.

% Participant observation allows us to carry out deep observation and reflection
% of the SOC's practices, and obtain thick data regarding the analysts' work.
% While such observation and analysis are about one particular SOC with its own
% unique characteristics, this limitation is compensated by the depth and thickness
% of data we were able to obtain which allowed for revealing patterns that
% are generalizable to other SOCs.

\subsection{AI companion design and deployment}

Alert triage emerged through fieldwork as the most
promising opportunity for intervention. Our observations % , informed by grounded
% theory~\cite{Strauss1990Sage} but not a full grounded-theory analysis,
showed that triage combined three characteristics: it was frequent, repetitive, and
strongly shaped by local organizational practice. Analysts repeatedly gathered
evidence from multiple systems, yet the final decision still depended on human
judgment informed by local knowledge and organizational context. Existing SOC
tools supported individual lookups but did little to reduce the repetitive work
of assembling evidence {\it across systems.}
These observations led us to focus the Companion on evidence gathering and
report drafting while leaving the final decision under analyst control.
The initial version of the Companion was first used by the two embedded researchers during
their own triage work. After several design iterations and when the researchers
were comfortable relying on its output to close tickets, the Companion
was introduced to other SOC analysts who already gained good SOC experience
(as determined by the SOC managers) for the final four months
of the study. Use was voluntary: analysts decided whether to use the
Companion on a ticket, whether to rely on its output, and whether to
continue the investigation manually.
Before deployment, analysts were briefed on the Companion's
capabilities, limitations, and available data sources, and were told
to treat its output as decision support rather than authoritative
judgment. Throughout deployment, the researchers remained available to
answer questions, observe use, and collect feedback.
% The companion was deployed in the SOC and first used by the
% two researchers themselves. After a few iterations and when
% the researchers themselves became comfortable using the companion's
% result to close triage tickets, 
% % After the initial design and development phase, the SOC AI Companion
% % was deployed in the same SOC where the fieldwork was
% % conducted.
% we invited other analysts to use the AI companion during live
% ticket work in the final four months of the study. Use was
% voluntary. Analysts could choose whether to use the companion on a
% given ticket, whether to rely on its report, whether to ask follow-up
% questions, whether to modify prompts, and whether to ignore the output
% and proceed manually as the tickets were still real and their own
% responsibility.

% Before deployment, the analysts in the SOC were kept in the loop
% consistently throughout the creation process explaining what the
% companion could do, what data it could access, and what its
% limitations were. They were told that generated outputs should be
% treated as assistance rather than authoritative decisions. The
% researchers remained available during the deployment period to answer
% questions, observe usage, and collect feedback.

% During the final four
% months, all SOC analysts were invited to use the developed AI companion during
% their regular work on live alerts. Use of the companion was voluntary,
% and analysts retained responsibility for reviewing any generated
% output before using it in a ticket.

\subsection{Data collection and analysis}
\label{sec:data_methodology}

% Researchers worked as part of the team while also collecting sociotechnical
% data reflected both in observation data and design implementation. Thus,
% we had the chance to observe differences in how analysts accomplished
% their daily tasks with and without the SOC Companion. Given our embedded
% approach, controlled experiments with and without the AI Companion were
% not feasible. Instead, data collected from our fieldwork reflect analysts’
% authentic interactions with the Companion in solving real SOC tasks.

As part of ensuring the trust we earned within the SOC, our embedded
research approach aimed to generate minimal disruption for analysts.
Controlled experiments were not feasible. Instead, data collected from
our fieldwork is observational in nature and reflect analysts’ authentic
interactions with the Companion in solving real SOC tasks.

We collected three sources of data: 1) field notes documenting
researchers' observations of SOC work, interactions with analysts,
and analyst feedback from interviews and informal comments; 2) Companion logs
containing prompts, AI model responses, tool activities, evidence, prompt versions, and
timestamps; and 3) ticketing-system records containing analysts' final
reports.
We analyzed these data using complementary qualitative and quantitative
approaches. % Two researchers independently coded Companion transcripts and final
% ticket reports to characterize how analysts used, modified, and responded to the
% Companion's output. We also measured draft reuse, verdict agreement, analyst
% interventions, and ticket completion time. Completion time was compared with
% analysts' self-reported estimates of manual investigation time, which we treat
% as approximate baselines rather than direct measurements.
Section~\ref{sec:analysis} presents the detailed analysis and findings.

%% file: design.tex
\section{SOC AI Companion Design Evolution}
\label{sec:design}
% -------------------------------------------------------------------------------

% ----------------------------------------------------------------------------------
%%%%%%%%%%%%% Moved from section 3 %%%%%%%%%%%%%%%%%%%%%%%%%%%%%%%%%%%%%%%%%%%%%%%%%
% ----------------------------------------------------------------------------------

\subsection{Design principles}

% The SOC AI Companion was built incrementally over the fieldwork.
% Our embedded researchers began by shadowing analysts on shifts;
% after this training they took shifts themselves on the same queues.

Working the alert queue like the other analysts allowed our researchers
to reflect upon the workflow and this reflection made the need for
AI Companion's features apparent.
% These requirements were shaped by the analysts in the
% SOC as a form of co-creation to fit analysts needs.
The Companion was not to
remove the analysts from triage, but to reduce repetitive
evidence-gathering and reporting labor while allowing analysts to
inspect, redirect, or revise the Companion's work.
Processing a typical alert required moving through
a number of internal and external systems to gather
relevant information.
Each system has its own interface, authentication, query conventions, and output format, and
none is aware of the others; the analyst was the integration layer: much of a shift
goes to copying a value out of one browser tab, pasting it into the next, and waiting,
while the judgment that goes into closing the ticket -- whether this is a real intrusion or another
benign event that tripped a rule -- is a small fraction of the elapsed time.
That asymmetry is what we targeted: not automating the decision but the assembly of
the context on which it rests.
Since analysts remained responsible for their
decision in closing tickets, %  the system could not simply provide a
% conclusion;
the Companion needed to
present evidence in a form that matches what the analyst would
produce manually. This not only eased adoption but also allowed
for verification by analysts. %  --  it needed to show the evidence used to reach that
% conclusion.
Our fieldwork also showed that analysts differed in their preferred
level of detail, report structure, and investigative habits. A useful
AI assistant therefore needed to be adaptable rather than behave
as a fixed, one-size-fits-all tool.

% Second, the Companion needed to be adaptable to local SOC
% practice. 

% Insights from the embedded fieldwork were translated into requirements
% for the AI companion. This subsection describes the methodological
% origin of the requirements; Section~\ref{sec:design} describes the
% system architecture and implementation.

% First, the Companion needed to begin from the ticket itself. Analysts
% should not have to manually copy alert details into a separate chatbot
% or reconstruct the relevant context from scratch. Second, the
% Companion needed access to the same types of evidence sources analysts
% used during triage, including log sources, enrichment services, asset
% information, and threat intelligence.

% ----------------------------------------------------------------------------------
%%%%%%%%%%%%% End %%%%%%%%%%%%%%%%%%%%%%%%%%%%%%%%%%%%%%%%%%%%%%%%%
% ----------------------------------------------------------------------------------

%\input{figures/soc_ticket_example}

%---------------------------
\begin{figure*}[t]
\centering
\includegraphics[width=.95\textwidth]{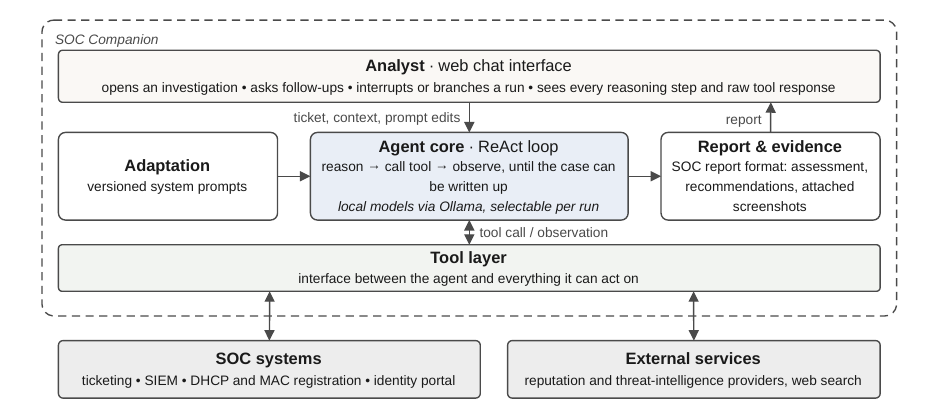}
\caption{\label{fig:architecture} System Architecture of SOC AI Companion}
\end{figure*}
%---------------------------

\subsection{Workflow automation}
\label{sec:workflow-ex}

 % The Companion did not close tickets automatically. Analysts
% remained responsible for reviewing the generated report, validating
% the evidence, and writing or approving the final ticket notes.
The key part of the Companion's work is to gather and assemble relevant information
from four kinds of systems:
1) several OSINT
(Open-Source Intelligence) and
threat intelligence services to establish whether the external indicator was
malicious; 2) the SIEM, to find which
internal host had communicated with it around the alert timestamp; 3) an internally
maintained DHCP lease lookup tool, to resolve the address the SIEM returns to a MAC
address and hostname; and 4) the device
identification portal, to attribute the device to a person.
This is a sequence of dependent lookups, where the next query follows from
what the last one returned, which is the pattern that ``tool-using agents''~\cite{Qin2024ACM,Yao2023ICLR} are suited to.
We therefore built the Companion as a ReAct-style agent~\cite{Yao2023ICLR} using
LangChain~\cite{langchain}, alternating between reasoning about what it still needs and calling a
tool to obtain it, until it can write up the case.
We ran it on locally hosted models, whose
reach is limited by the tool layer: % it executes
% no code on the host \xo{which host? Surely some code gets run on the Spark machine?}
% and
it cannot compose actions of its own but only calls from the fixed
set of functions preconfigured in the system, with schema-validated arguments. All the tool calls are
read-only -- retrieving a ticket, running a hunting query, looking up a reputation
record, and so on -- and cannot modify any state in the SOC's systems. A run that goes wrong
therefore costs a wasted investigation, not a change to the systems the SOC depends on.
Most importantly, no sensitive information
such as internal addressing, hostnames, or the identity of a community member leaves
the organization's network.

\subsection{Empowering analysts to adapt the AI}
\label{sec:adapting}

% Working with different analysts and shadowing their investigations yielded a second
% insight. 
Analysts who close the same alert type reach the same conclusion by different
routes: some enrich every external indicator before touching internal logs; others
identify the host first and drop the enrichment once the traffic admits a benign
explanation, such as a routine update service.
Their write-ups differ too, in ordering and in how much raw log data and ruled-out
indicators they retain.
An AI agent with a fixed procedure would have suited some and irritated the
others; so we made the Companion's behavior itself something analysts could direct.
At this time the primary adaptation mechanism is through the system prompt, which
describes the workflow, the order in which
tools are tried, the conditions for abandoning a line of inquiry, and the report
layout.
Analysts kept several named versions of system prompts and switched between them per investigation,
with a shared base version as the default for anyone who has not written their own.
By adapting system prompts the analysts can influence how the Companion works
by themselves without relying on the researchers or redeployment.

\subsection{Deployment and refinement}
\label{sec:initial_deployment}

We deployed the Companion on an NVIDIA DGX Spark~\cite{dgxspark} inside the SOC and gave accounts
to a small group of experienced analysts. Their feedback led to two refinements 
during the deployment. The first extended the tool layer to capture visual evidence:
whenever a tool performs a lookup, a headless browser driven by
Playwright~\cite{playwright} also visits the corresponding page,
renders the HTML the service returns, and captures a screenshot of the result.
The second refinement added a revised version of the shared default system prompt,
which changed only the report layout based on analysts' feedback 
without changing the reasoning or tool usage.
Section~\ref{sec:analysis} reports the impact of both refinements.

\subsection{Final design}
\label{sec:design:final}

The design that emerged from this process has four parts, shown in
Figure~\ref{fig:architecture}.

\noindent\textbf{Agent core.}
A ReAct-style agent loop drives each investigation.
The analyst opens an investigation by supplying a ticket number or pasting an
alert, and the agent retrieves the full ticket
(Figure~\ref{fig:appendix-alert} in Appendix~\ref{sec:appendix-example} shows
one), then iterates between reasoning and tool calls until it can produce a
report.
Runs execute on the server rather than in the browser, so an investigation
survives a page refresh or a closed laptop, and a run that exceeds its time budget
is terminated and reported as failed rather than left hanging.

\noindent\textbf{Tool layer.}
The tools available to the model fall into the categories shown in
Table~\ref{tab:tools}, which together cover the systems an analyst navigates
manually during triage.

%---------------------------
\begin{table}[t]
\small
\begin{tabular}{@{}p{0.26\columnwidth}p{0.69\columnwidth}@{}}
\hline
\textbf{Category} & \textbf{Systems reached} \\
\hline
Ticketing & Alert ticket retrieval \\
Log sources & SIEM advanced hunting queries, log search platform \\
Internal asset and identity & DHCP lease lookup, MAC registration, device and owner records \\
External enrichment & Reputation and threat intelligence providers, general web search \\
Analysis utilities & Encoding and decoding, detection signature lookup \\
\hline
\end{tabular}
\caption{\label{tab:tools} Categories of tools available to the Companion.
  Each category corresponds to systems the analysts already use manually
  during triage.}
\end{table}
%---------------------------

\noindent\textbf{Adaptation layer.}
Editable and versioned system prompts allow
the Companion's behavior to be changed by its users without our involvement.

\noindent\textbf{Report and evidence layer.}
The agent's output is a structured report following the SOC's own reporting
conventions, with an assessment, recommendations, and a supporting evidence
section carrying the screenshots captured during the investigation.

% Analysts interacted with the Companion through a 
% interface that allowed them to import a ticket.  The Companion may pause asking for additional information
% from the analysts.

Analysts work with the Companion through a web-based interface.
After a ticket is imported
a series of actions happen automatically based on the
AI reasoning.
Every reasoning step, tool invocation, and raw tool response is streamed to the
interface as the run proceeds and retained afterwards, so an analyst can see which
query produced a given claim. %rather than being asked to accept the summary.
In the end the Companion generates a structured
triage report in a similar form as what analysts would produce
themselves. Analysts inspect the report; they can provide additional context,
request further investigation, ask follow-up
questions about the case, branch a conversation to explore an alternative
reading of the evidence, or decide to modify system prompts and other settings
and rerun the investigation. 
The interface also records everything we later analyze in
Section~\ref{sec:analysis}: transcripts, tool calls, prompt edits, and the timing of each run.
Inference runs on models served locally on the DGX Spark with Ollama~\cite{ollama}, and the model
used for an investigation is selectable per run, which let us compare locally hosted
models of different sizes during the deployment without changing anything else about
the system.
We used GLM-4.7-Flash (30B)~\cite{glm47flash} for most of the deployment, since it called tools
reliably and was small enough to run alongside the rest of the system on a single
node.
Qwen3.6 (35B)~\cite{qwen36} and gpt-oss (20B)~\cite{Agarwal2025Arxiv} were available too, and an analyst could switch to
either of them for a given investigation.

%%% Local Variables:
%%% mode: latex
%%% TeX-master: "main"
%%% End:

%% file: data-analysis.tex
%-------------------------------------------------------------------------------
\section{Data Analysis and Findings}
\label{sec:analysis}
%-------------------------------------------------------------------------------

An analyst closes a SOC ticket by writing a report and attaching it to the
ticket. Whatever the AI companion produces is therefore useful only insofar as it
reaches that report, and this is a question our deployment can answer directly.
% rather than by asking analysts what they thought of the tool.
The Companion
retains every thread, message, tool call, and raw tool response, along with the
system prompt and model that produced each run. The ticketing system retains
what the analyst filed at the end of it. Setting the two records side by side
shows what the Companion offered and what the analyst was willing to put their
name to, for the same ticket.

% We acknowledge that an analyst being willing to use the Companion's result
% does not automatically mean the result is correct. In an ideal world
% an experiment could have been carried out where another analyst (or multiple of them)
% redo a ticket investigation without using the Companion at all and researchers compare the
% results. Such an experiment however would disrupt the SOC's normal work and would
% have certainly cost us the trust and good will we earned that enabled us to
% conduct the fieldwork in the first place. On the other hand, since the adoption of
% Companion's result was done by analysts in actual work with real consequences
% -- they would be blamed by closing a ticket wrong -- the quality of analysts'
% judgment in adopting the Companion's result shall not be undervalued. The SOC
% manager periodically checked all analysts' work; more senior analysts were also
% asked to review other analysts' work. These real-world quality control methods
% are mitigation for the imperfection in analysts' judgment and using it as a proxy
% to estimate the Companion's usefulness.

\subsection{Corpus and coding}
\label{sec:analysis:corpus}

Our data include all tickets that SOC analysts, excluding the embedded
researchers, ran through the Companion during the last four months in the
fieldwork starting on April 29, 2026. For coding and more in-depth analysis, we cut-off
on the ticket date of July 31, 2026 with a total number of 108 tickets.
% Our data include all tickets the SOC analysts (not including our researchers)
% ran the Companion on between April 28 and July 31, 2026.
% The total number of
% these tickets is 108.
There were 10 analysts working in the SOC during that time period and
we only made the Companion available to analysts who were determined by
the SOC managers as more experienced, to reduce the likelihood that analysts miss
incorrect information in the Companion's output.
These analysts had all worked in the SOC for at least six months and many had
already worked there for a year.
% had already worked in
% the SOC for at least six months, 
% \xo{How many analysts fall into the ``more experienced'' group that we made
%   the Companion available? All six?}
Six analysts (referred to as A1--A6) appeared in the corpus.
A2 contributed 42 tickets, A3 contributed 39, A5 contributed 19, A4 contributed 4,
A6 contributed 3, and A1 contributed 1.
Not all six analysts were present for the entire
deployment period. A1 left the SOC a week after the Companion became available
to them, A4 left partway through the period, and A6 received access to the Companion only in
its final two weeks. A2, A3, and A5 were present for most of the deployment.
94 tickets were worked in a single Companion thread; 14 were worked across two or more.

Our data corpus aims at capturing whether and how analysts used the Companion's output
in closing the tickets. To this end
we associate each ticket with two records: the Companion transcript and the final
ticketing-system report. The transcript included messages, tool calls, raw tool
responses, generated screenshots, prompt versions, model selections, and
timestamps. The ticketing-system report contained the text and images the
analyst ultimately filed and signed their name to. We coded each pair using
three sets of codes. Usage codes (U1--U14) describe what happened to the
Companion's draft, including whether its text or screenshots reached the final
report and what was added, removed, or reversed. Intervention codes (I1--I11)
capture analyst messages sent after the opening request that started the run.
Other codes (O1--O5) capture reruns, prompt or model changes, and grounding
failures such as fabricated assertions or invalid tool calls. Table~\ref{tab:codebook}
lists all codes and frequencies. We also recorded the verdict each side reached,
grouping both the Companion's draft verdict and the analyst's final verdict into
three classes: \emph{benign}, \emph{uncertain}, and \emph{malicious}.

\begin{figure}[!t]
    \centering
    \includegraphics[width=.95\columnwidth]{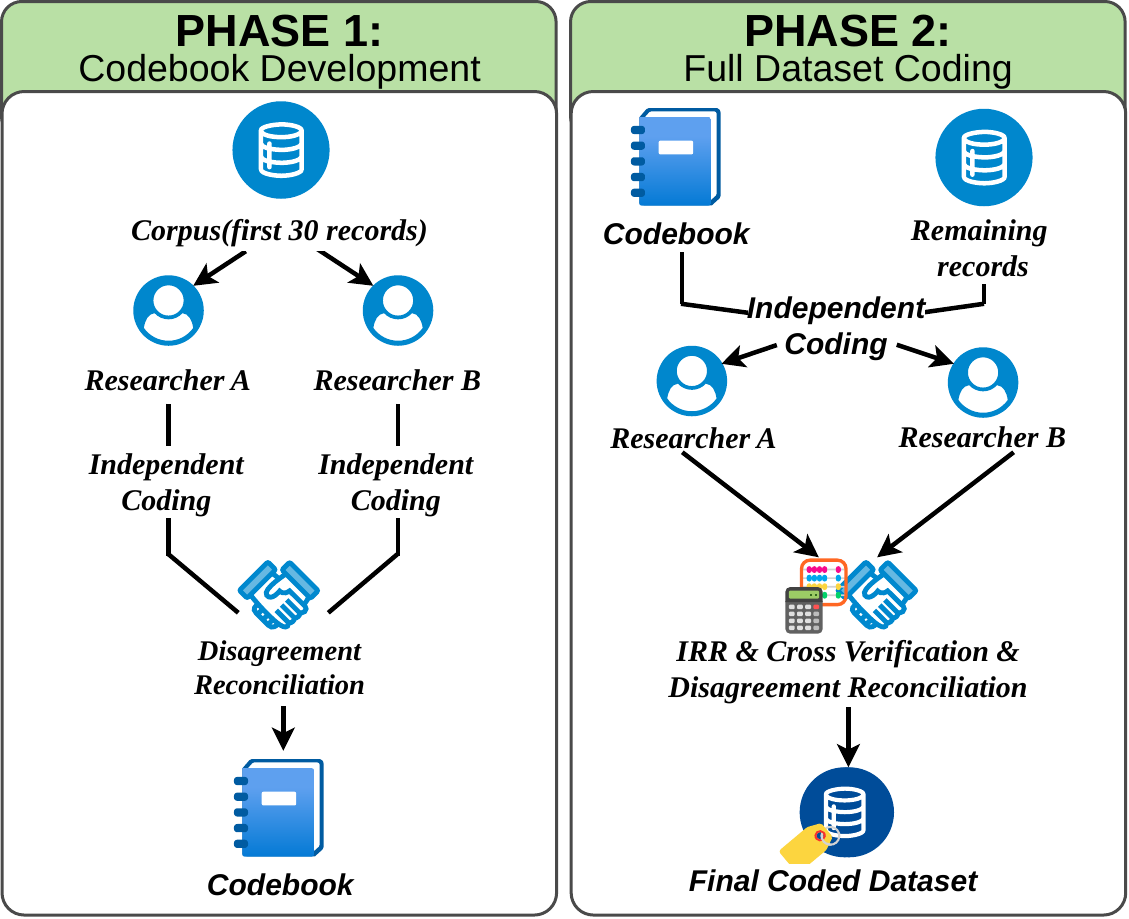}
    \caption{Coding Process.}
    \label{fig:coding}
\end{figure}

The two researchers first independently coded an initial 30 tickets'
records, compared their assignments, and reconciled disagreements to
refine and finalize the codebook. Using the resulting codebook, both
researchers then independently coded the remaining 78 tickets' records.
Figure~\ref{fig:coding} illustrates this process.
We assessed inter-rater reliability on these independent assignments
using Cohen's $\kappa$~\cite{Cohen1960ACO}. Across 2,340 binary coding
decisions, the researchers agreed on 2,322, yielding 99.23\% observed
agreement and an overall Cohen's $\kappa=0.965$. Following the
reliability assessment, the researchers reconciled the remaining
disagreements to produce the final coded dataset.

%---------------------------
\begin{table}[t]
\small
\begin{tabular}{@{}p{0.10\columnwidth}p{0.62\columnwidth}r@{}}
\hline
& \textbf{Usage codes} & $n$ \\
\hline
U1 & Text used verbatim & 35 \\
U2 & Text partially used & 63 \\
U3 & Text not used & 10 \\
U4 & Analyst added text & 26 \\
U5 & Text truncated & 41 \\
U6 & All screenshots reused & 78 \\
U7 & Some screenshots reused & 17 \\
U8 & No screenshots reused & 10 \\
U9 & Analyst added own screenshots & 51 \\
U10 & Recommendation overridden to no-action & 34 \\
U11 & AI evidence gap filled by hand & 12 \\
U12 & AI verdict uncertainty suppressed & 26 \\
U13 & AI template artifact cleaned & 5 \\
U14 & AI verdict escalated & 2 \\
\hline
& \textbf{Intervention codes} & \\
\hline
I2 & Evidence demand & 6 \\
I6 & Output shaping & 5 \\
I1 & Missed step & 4 \\
I5 & Provide data & 4 \\
I3 & Error diagnosis & 3 \\
I4 & Provide context & 3 \\
I8 & Widen search & 2 \\
I7 & Verdict steering & 1 \\
I9 & Nudge & 1 \\
I10 & Off-task & 1 \\
I11 & Ask for explanation & 1 \\
\hline
& \textbf{Other codes} & \\
\hline
O1 & Ticket rerun in a new thread & 14 \\
O2 & Different system prompt tried & 9 \\
O3 & Different model tried & 1 \\
O4 & AI hallucination or fabrication & 1 \\
O5 & Invalid tool invocation & 2 \\
\hline
\end{tabular}
\caption{\label{tab:codebook} Codebook and code frequencies over the 108 coded
  tickets. Except for U1--U3 and U6--U8, which are each mutually exclusive,
  codes may co-occur; a ticket carries as many as apply. Definitions are given
  in Appendix~\ref{sec:appendix-codebook}.}
\end{table}
%---------------------------

Alongside the codes, we compute \emph{draft reuse}: a text-only similarity
score between the Companion's draft report and the analyst's final closing
report.
% Alongside the
% codes we compute \emph{draft reuse}, the textual similarity between the
% Companion's draft and the analyst's closing report.
% Draft reuse is computed per ticket over text only.
For each ticket, on the Companion side we take
the draft report it produced in the primary thread; on the analyst's side we take
the closing report they filed on the corresponding ticket.
We normalize both by removing
formatting markers, collapsing whitespace, and folding case, then score the
pair using RapidFuzz's token set ratio~\cite{rapidfuzz}, a word-order-insensitive
variant of edit similarity~\cite{Levenshtein1966}.
% formatting markers dropped, whitespace collapsed, case folded --- and score the
% pair with the token set ratio of RapidFuzz~\cite{rapidfuzz}, a
% word-order-insensitive normalization of edit similarity~\cite{Levenshtein1966}.
The comparison is set-based: the two texts are compared as bags
of words, making the score insensitive to reordering and to reflowed
formatting. % Screenshots carry no tokens and are excluded from the score; whether
% the analyst kept the Companion's images is tracked separately by codes U6--U8.

% The resulting score cannot be interpreted as the fraction of the draft's words that survived.
% A good way to interpret the draft reuse scores is to look at their correlation with the codes,
% which were assigned by our researchers reading the two reports:
% The 35 tickets coded U1
% (text used verbatim) score a median of 99.6\% (IQR 98.8--100.0); the 10 coded U3
% (text not used) score a median of 51.3\% (IQR 49.0--74.2).

Tickets coded U1
(text used verbatim) had a median reuse score of 99.6\% (IQR 98.8--100.0),
whereas tickets coded U3 (text not used) had a median of 51.3\%
(IQR 49.0--74.2). 
A U1 case does not always have 100\% reuse score
% reads at roughly 95\% and above rather  than 100\%, 
because the two documents
may differ in non-substantive ways; for example
% each document contains something the other lacks, and here neither
% difference is substantive
the closing report leaves out the Companion's opening
line (``Now I have all the necessary information. Let me generate the final
report.'') and carries captions for the attached screenshots (``DHCP Lease'',
``Recorded Future IP'').
On the other hand, a closing report the analyst wrote from scratch
still has roughly 50\% reuse score rather than zero, because it still talks about the same
incident and thus shares many common elements --- addresses, hostnames, and
signatures with the Companion report.

Some properties of the corpus limit the conclusions one can draw.
First, analysts used the AI Companion and were influenced
by its reasoning. Thus they were not independent evaluators for accuracy.
This research focused on how the Companion was utilized; next steps
could include more independent assessments of the Companion’s accuracy
(e.g., via hiring independent analysts to redo some tickets). However,
even outside assessments won’t necessarily capture analysts’ judgment
on whether/how to use the Companion –- these can be best assessed
through what happens in practice.

Second, use of the Companion was voluntary and analysts chose which alerts to bring to it; thus the
corpus is not a random sample of the alert queue. The analysts are unevenly
represented, so we make no claim resting on A1, A4 or A6 alone. The
Companion changed during the deployment --- its system prompt was revised and
its model swapped --- so the corpus is not homogeneous over time.
In this case, the combination of participant observation and design iteration
is precisely what can capture analysts’ authentic behaviors while using the
Companion in the SOC’s real tasks. % with real-world consequences.

% Despite these limitations, our data corpus captures analysts' authentic behaviors
%   while using the Companion in the SOC's real tasks with real-world consequences.
%   The rationale of using this
% observational approach in data collection was discussed in Section~\ref{sec:data_methodology}.

% Second, use of the Companion was voluntary and analysts chose which
% alerts to bring to it; thus the corpus is not a random sample of
% the alert queue. The analysts are unevenly represented, so we make
% no claim resting on A1, A4 or A6 alone. Moreover, the Companion
% changed during the deployment — its system prompt was revised and its model swapped— so the corpus is not homogeneous over time. 

% Some properties of the corpus limit the conclusions one can draw from it.
% First, analysts in our corpus used AI Companion and thus are {\it not} independent
% evaluators of the Companion's result. For this reason our data cannot yield an independent assessment
% of the Companion's accuracy. Rather it aims at providing insights into analysts'
% judgment on whether/how to use the Companion's output and those judgments are certainly
% influenced by the Companion's use.

\subsection{How much of the Companion's work reached the ticket closing report}
\label{sec:analysis:reach}

\begin{figure}[t]
\centering
\includegraphics[width=\columnwidth]{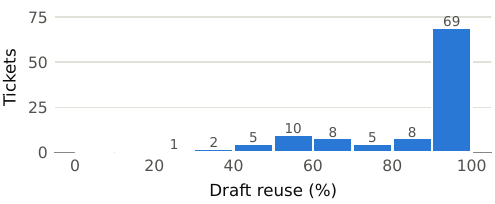}
\caption{\label{fig:usage-hist} Share of the Companion's draft surviving into
  the analyst's closing report, over 108 tickets. Drafts are taken nearly whole
  or discarded, with little in between: 59 tickets at or above 95\% reuse, 28
  below 75\%. Median 96.7\%, mean 85.5\%, lower quartile 72.0\%.}
\end{figure}

Across the corpus the median draft reuse is 96.7\%, the mean 85.5\%, the lower
quartile 72.0\%, and the minimum 27.5\%. The median alone would suggest that
analysts accept drafts almost entirely. But the average differs by eleven
points. Figure~\ref{fig:usage-hist} shows why: the distribution is highly concentrated
towards high reuse rate (close to 100\%), with a second, smaller concentration of
low-reuse tickets (around 50\%). Recall that a 50\% reuse score indicates the
analyst rewrote the report from scratch.
% has two
% ends, and neither average describes it.
The median sits completely inside the high-reuse
group (90-100\%), which holds 69 of the 108 tickets, and the mean is dragged by the lower
reuse cases. % The size of each end is the
% real summary:
Of the 108 tickets, 59 (55\% of them) have at least 95\% draft reuse and
28 (26\% of them) have less than 75\% draft reuse; the remaining 21 fall in between. The codes tell
the same story: 35 tickets were closed with the draft text reused verbatim (U1) and 10 were
written from scratch (U3). The 63 partial-use tickets (U2) lean towards the two
ends --- 25 at or above 95\% and 21 below 75\% --- rather than filling the middle.

Analysts, in other words, do not edit drafts often. They decide whether to
take one and that decision is close to binary. When a draft is taken it is
often taken whole, and when it is rejected the analyst writes the closing
report from scratch rather than taking parts of it that were sound.
Table~\ref{tab:partial} breaks down the partially-used Companion reports (code U2) by how they were
edited. The most common edit is not to the analysis at all. In 20 tickets the
analyst pasted the draft essentially whole --- median reuse 97.9\%, mean 92.2\%
--- and changed only its recommendation. We return to those in
Section~\ref{sec:analysis:layers}.

%---------------------------
%---------------------------

%---------------------------
\begin{table}[t]
\small
\begin{tabular}{@{}p{0.40\columnwidth}rp{0.20\columnwidth}p{0.20\columnwidth}@{}}
\hline
\textbf{How the AI draft was edited} & $n$ & \textbf{Mean reuse score} & \textbf{Median reuse score} \\
\hline
Recommendation dropped & 20 & 92.2\% & 97.9\% \\
Trimmed & 8 & 92.8\% & 96.2\% \\
Verdict softened & 8 & 83.2\% & 93.8\% \\
Condensed & 7 & 68.3\% & 64.6\% \\
Rewritten & 7 & 61.5\% & 59.7\% \\
Reframed & 6 & 68.5\% & 67.1\% \\
\hline
Kept and extended & 4 & \multicolumn{2}{c}{61.5\%, 82.3\%, 85.8\%, 90.7\%} \\
Cleaned paste & 3 & \multicolumn{2}{c}{91.5\%, 97.4\%, 99.6\%} \\
\hline
\textbf{Total} & \textbf{63} & \textbf{82.2\%} & \textbf{91.5\%} \\
\hline
\end{tabular}
\caption{\label{tab:partial} Partially-used Companion drafts (U2) by type of edit, with mean
  and median draft reuse. The most common edit leaves the analysis intact and
  removes only the recommended action. For the two categories with $n<5$ we
  list the individual ticket's draft reuse values rather than an average.}
\end{table}
%---------------------------

\subsection{Verifiability is key to adoption}
\label{sec:analysis:verifiability}

% We observed that even though the Companion's work remained the same, analysts were
% more eager to use it after we changed how it presented its findings.
Analysts told us % the Companion report drafts
% described the evidence instead of showing it in a non-disputable way. In particular,
% analysts
they would always attach the screenshots from the various information sources (Virus Total, SIEM,
etc.) in their report, while the initial version of the Companion presented API query results from
those tools in a textual format. Even though the textual content was correct information, analysts
were not willing to accept them. One analyst even insisted on verifying the results
manually by querying each tool themselves. This certainly defeats the purpose of
reducing analysts' burden. 
We realized that for analysts to adopt the AI Companion, it is not sufficient to
solely ensure all information presented by the Companion is accurate; it
must also show the accuracy in a self-evident way. Incorporating visual rendering
of the result from the external tool serves that purpose --- it becomes clear that
the evidence was not made up, but from a trusted source. In fact this is the exact
reason why analysts included screenshots in their own reports.
We then extended the Companion's tool layer to render and capture the page behind
each tool lookup and attached the ``screenshots'' to the draft, as described in
Section~\ref{sec:initial_deployment}.

The first ticket carrying attached screenshots
was on May 28, 2026. There were only three tickets
predating that despite the Companion having been made
available to analysts for more than four weeks at that
point. After the screenshot feature was introduced,
the researchers noticed an obvious change of analysts'
attitude towards the Companion -- they became much more
willing to use it. In fact this is the moment when we saw
the largest shift in adoption during the deployment.
Draft reuse reflects this change: % rose sharply over the same period:
across the three tickets before the change
the median reuse was 50.0\% %  and one of the three was kept nearly whole
% ($\ge 95\%$ reuse),
whereas across the 105 tickets after it the median was
96.7\%. % and 58 (55\%) were kept nearly whole.
The mean rises more modestly, from
66.0\% to 86.1\%.
Because only three tickets predated the change, we treat the before/after
reuse comparison as descriptive rather than a controlled baseline.
% a few
% drafts were still rewritten from scratch, and those pull the average down while
% the majority of the Companion drafts were kept nearly intact.
Table~\ref{tab:visual-evidence} illustrates this statistics.

% that and 105 followed it.
% After this visual evidence layer was added, the Companion produced the same
% investigation narrative as before; what changed is that the narrative now arrived with the
% % page behind them
% screenshots of tool results (not just the text describing the results)
% attached.
% This is
% The
% clearest sign of it is how much of the SOC's ticket queue the analysts chose
% to bring to the Companion.

\begin{table}[t]
\small
\begin{tabular}{@{}lrrrr@{}}
\hline
& $n$ & \textbf{Mean} & \textbf{Median} & $\ge 95\%$ \\
& & \textbf{reuse} & \textbf{reuse} & \\
\hline
Before & 3 & 66.0\% & 50.0\% & 1 (33\%) \\
After & 105 & 86.1\% & 96.7\% & 58 (55\%) \\
\hline
\end{tabular}
\caption{\label{tab:visual-evidence} Comparison before and after screenshots were attached
  to the Companion's draft reports.}
\end{table}

% Since only three tickets predated this change,
% the ``Before'' row is more an anecdote than a baseline, and we read the
% comparison as suggestive on its own. The usage codes carry more weight.
The usage codes provide another view on how analysts treated the visual
evidence. Of the 105
tickets after the change, 78 (74\%) carry U6 (all screenshots reused),
and a further 17 (16\%) carry U7 (some screenshots reused) --- 90\% of tickets kept at
least some of the screenshots the Companion captured.
This indicates the importance of the screenshots to the analysts.
% Neither code appears before the change, for the trivial reason that there was no screenshot to reuse.

% We take this as evidence that what governed adoption was not only
% whether the Companion's analysis was accurate but also whether its output took a form
% the analysts could quickly verify.
This experience seems to indicate that adoption depended not only on whether the
Companion's analysis was accurate, but also on whether its output was quickly
verifiable.
By \emph{verifiability} we mean whether an analyst
could confirm the information by themselves without redoing the work. 
% A draft report that is accurate may be of little use 
% if the analyst has to repeat every lookup to be sure of it so they can sign their name on it.
A model-written sentence that an
indicator was flagged by a reputation service is still a claim the analyst must
check. A screenshot of that reputation-service page, captured by the tool layer
from the authoritative source, gives the analyst evidence they can inspect and
attach to the ticket.
% An assertion that an indicator is flagged
% by a reputation service is a claim the analyst must independently verify before
% filing it; a capture of that service provider's page showing the flag serves as the
% verification in itself,
It is also what a reviewer of the report expects to find
there. The difference is where the two come from. Text is written by the model,
and a model can make up text so an analyst cannot tell a true result from API call
% summary of a page
from a hallucinated one without calling the tool at all. 
A screenshot is produced by the tool layer and the content comes
from the authoritative source's page.
% itself, so the
% image can only have come from that source.
The model writes the text, but it
cannot generate the image itself.\footnote{The model used was a text-only LLM.}
The Companion's usefulness increased after the change even though its
substantive work stayed the same.

% It should be acknowledged that a more powerful
% model than the one we used could have hallucinated images as well, and one can
% even imagine that a ``rogue model'' could choose to not attach the image from
% the authoratative tools but make up one by itself. This poses an interesting
% question for future research --- how could human analysts cultivate the right
% level of trust towards an AI companion. What mechanisms could ``incentivize'' an AI agent
% to not lie? It is beyond the scope of this paper to address this question.
% In our AI SOC companion each attached screenshot is associated with an URL
% with all the query parameters populated. A suspecting analyst could always
% click the URL and go to the authoratative source themselves to manually verify.
% In fact in our fieldwork we found one analyst who often performed this second
% verification. 

% By \emph{accuracy} we mean whether the Companion's analysis was right: whether it
% queried the correct indicators, read them correctly, and reached the conclusion
% an analyst would have reached. 

\subsection{Three layers of adoption: content, verdict, and recommendation}
\label{sec:analysis:layers}

Draft reuse measures whether the Companion's prose reached the ticket.
It does not measure whether its judgment did, and the two diverge sharply.

\begin{figure}[t]
\centering
\includegraphics[width=.95\columnwidth]{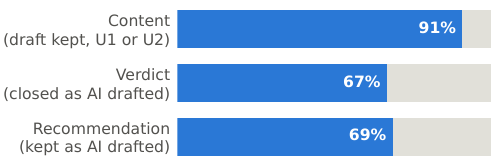}
\caption{\label{fig:layers} Three levels at which the Companion's output could be
  adopted: content, verdict, and recommendation. Each bar is the share of the same 108
  tickets on which that layer survived into the closing report.}
\end{figure}

% \xo{I noticed there is no code for these. Did you do this analysis manually
%   each time?}
We put three questions to the same corpus, each answered as a share of the same
108 tickets. Did the analyst keep the Companion's content? On 98 tickets, or 91\% of
all,
the closing report reused the draft text whole (U1) or in part (U2).
Did the analyst agree with the
verdict the Companion reached? In 72 of 108 cases, or 67\% of all, the answer is yes.
% comparing the two
% verdicts as the three classes of Section~\ref{sec:analysis:corpus}. 
Did the
analyst adopt the recommendation the Companion produced? Every draft ends by
recommending what to do about the alert, which may be to do nothing.
In 74 of them, or 69\% of all, the analyst kept the recommendation
made by the Companion (some wording may have been shortened/condensed).
On 34
tickets --- roughly a third of the corpus --- the analyst deleted the action the
Companion had proposed and filed the report recommending no action (U10).
% Recommendation overrides run in this direction only:
We saw no case where the analyst added or
escalated an action the Companion had not recommended.
Figure~\ref{fig:layers} puts the three rates side by side.
The first rate stands well apart from the other two. The Companion's content
reached the closing report on 91\% of tickets, but its verdict did so on 67\%
and its recommended action on 69\% ---
the content reaches the closing report much more often than the judgment does.

%---------------------------
%---------------------------

The gap between the first answer and the third is telling. Analysts who
struck the recommendation had reused nearly as much of the
draft as analysts who kept it, on both statistics
(Table~\ref{tab:override-reuse}): a median of 94.2\% against 97.2\%, a mean of
83.4\% against 86.5\%.
Both groups sit high in the reuse distribution,
and the gap between them --- within three percentage points of median and mean --- is far
smaller than the gap between a kept draft and a rewritten one. Striking down the
Companion's recommendation
is therefore not telling us the analyst distrusted
the draft and rewrote it. Instead the analyst kept the Companion's investigation, 
its analysis and evidence, filed all of it, and cut the final sentences that 
recommended some actions. Keeping the content and accepting the judgment are separate acts,
and draft reuse score only reflects the first.

%---------------------------
\begin{table}[t]
\small
\begin{tabular}{@{}lrrr@{}}
\hline
\textbf{Companion's recommendation} & $n$ & \textbf{Median} & \textbf{Mean} \\
  \textbf{was} &  & \textbf{reuse} & \textbf{reuse} \\
\hline
struck down (U10) & 34 & 94.2\% & 83.4\% \\
kept & 74 & 97.2\% & 86.5\% \\
\hline
\end{tabular}
\caption{\label{tab:override-reuse} Draft reuse by whether the analyst kept or
  struck down the Companion's recommended action.}
\end{table}

%---------------------------

Where analysts struck the recommendation is just as revealing.
Table~\ref{tab:override} splits the tickets by the verdict the Companion had
reached and compares recommendation adoption.
Analysts never struck recommendation from a benign finding: a benign
finding recommends nothing, so there is nothing to strike. But they struck recommendations
from 24 of its 40 uncertain findings and 10 of its 30 malicious findings. For the malicious
U10 cases
the analysts also filed a lower verdict (uncertain or benign) than the Companion had drafted,
except in two cases. In these two cases the recommendation was struck because
the action had already been taken --- the user was already disabled, or the traffic
was already dropped. Hedging costs the
Companion its recommendation --- where it declined to commit, analysts kept its
work and made the call themselves. Code U12 (Table~\ref{tab:codebook}) points the same way: on 26 tickets,
roughly a quarter of the corpus, the analyst deleted the Companion's stated uncertainty
before filing. These findings indicate that the AI companion still provided
valuable utility to the analysts even when it could not reach as decisive
a decision as the analyst.

%---------------------------
\begin{table}[t]
\small
\begin{tabular}{@{}lp{1in}p{.8in}c@{}}
\hline
\textbf{Companion} & \textbf{Recommendation} & \textbf{Recommendation} & \textbf{\% U10} \\
\textbf{verdict} & \textbf{struck down (U10)} & \textbf{kept} & \\
\hline
Benign & 0 & 38 & 0\% \\
Uncertain & 24 & 16 & 60\% \\
Malicious & 10 & 20 & 33\% \\
\hline
\end{tabular}
\caption{\label{tab:override} Tickets where analysts struck down/kept
  recommendation by Companion verdict.}
\end{table}
%---------------------------

%---------------------------
\begin{figure}[t]
\centering
\includegraphics[width=0.92\columnwidth]{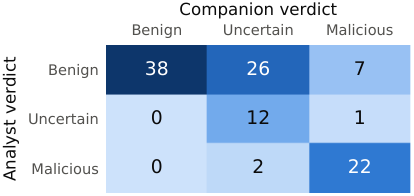}
\caption{\label{fig:confusion} The verdict analysts closed tickets as
  (rows) vs. the Companion's draft opinions (columns), over the 108 tickets. The Companion
  over-hedges: most of what it called uncertain was closed benign.}
\end{figure}
%---------------------------

Figure~\ref{fig:confusion} lays out the full verdict comparison; it is important
to note that this is not a confusion matrix despite displayed in a similar
form. As can be seen in the discussion that follows, verdicts at times reflect more
choices of judgment than absolute ground truth.
The Companion hedges more than analysts: of the 40 alerts it called
uncertain, analysts closed 26 as benign and left only 12 uncertain. Of the 26 benign cases,
15 were the same signature: a default Cloudflare tunnel domain pattern in the
TLS SNI\footnote{Server Name Indication, the TLS handshake extension that
  carries the destination hostname in cleartext, making it visible to network
  monitoring even for encrypted connections.}. Each alert looks anomalous on its own, so the
Companion declined to commit. But this is a university network where students
run their own devices, and a student standing up a tunnel is ordinary. The fact
that settles the alert is a base rate that lives in the analyst's head, not in
any tool the Companion can call. Hedging here is a missing prior, not a
reasoning failure.

Conflicting verdicts, not hedging, is what costs the Companion its draft.
When the two verdicts conflict outright, median reuse falls to 67.9\% and mean
reuse to 67.2\%, against 97.4\% and 87.5\% when the verdicts match
(Table~\ref{tab:verdict-reuse}). When only one side declined to commit (uncertain), reuse
barely changes --- a median of 96.4\% and a mean of 85.1\%.

For seven tickets the AI companion and analysts disagree outright on the verdict
(malicious vs. benign). In all those cases
the Companion drafted malicious and the analyst closed the ticket as benign.
Out of the seven only one was a clear Companion error. The alert reported that 
a Windows host made a TLS request to a known benign public IP-lookup service; 
a signature fired on that pattern because the malware infostealers uses such services 
to report a victim's address. The alert's reference block linked a VirusTotal 
page for one such stealer's hash value as documentation.
Nothing tied that hash to this Windows host.
The Companion draft also stated that no matching process events
were found on the host either. 
Despite this, the Companion 
still reported 
the malware sample as though it were found on this host, naming the malware and
writing that the traffic matched its behavior. 
It even looked up the hash in VirusTotal regardless.
This is a clear case of the Companion's confusion in understanding the SOC
ticket.
Besides this one case, the other six are judgment calls that could be decided either way.
% and they score
% against the Companion only because we treat the analyst's verdict as ground truth.
Five of them hinged on a single question --- how much a clean IP reputation should
count against a malicious signature? Four of those cases involved inbound traffic
that was already dropped by the firewall due to payload matching C\&C signatures.
The source sat in well known cloud provider space (AWS, Google, Azure).
% the destination sat in
% well known cloud provider space (AWS, Google, Azure).
The Companion's draft reported the source as
unflagged but decided malicious anyway.  The analyst
read the same lines the other way: since the traffic was already dropped
they closed the tickets as benign.
%without introducing any new evidence they

%---------------------------
\begin{table}[t]
\small
\begin{tabular}{@{}lrrr@{}}
\hline
\textbf{Verdict relationship} & $n$ & \textbf{Median} & \textbf{Mean} \\
 & & \textbf{reuse} & \textbf{reuse} \\
\hline
Verdicts match & 72 & 97.4\% & 87.5\% \\
One side hedged & 29 & 96.4\% & 85.1\% \\
Verdicts conflict & 7 & 67.9\% & 67.2\% \\
\hline
\end{tabular}
\caption{\label{tab:verdict-reuse} Draft reuse by how the analyst's verdict
  compares to the Companion's.}
\end{table}
%---------------------------

The transcripts record failures the verdict comparison could not show. On 12 tickets
the analyst supplied device, user, or payload attribution by hand because the
Companion's tools had come back empty (U11). On five it leaked its own prompt
scaffolding into the draft, and the analyst stripped it before filing (U13). On
three its own output was not grounded in its context. Twice it made an invalid
tool call (O5) --- an IP passed to a reputation tool that takes a domain, and a
signature-lookup tool called under a name it did not have --- and both failed
visibly at the tool layer. The third was silent: on a ticket whose alert was
itself a tunnel-domain-pattern signature, the Companion named a specific
subdomain matching that pattern as a suspicious indicator, one that appeared
neither in the alert nor in any tool result (O4). It was fabricated in the shape
the ticket had primed, and nothing in the draft set it apart from an indicator a
lookup had returned; the analyst caught it only by asking where the domain had
come from.

% Those two classes differ in kind.
% When an invalid tool call occurs, the analyst sees an error. A fabrication looks
% exactly like a result, and that difference in visibility is what most directly
% justifies the verification behavior we describe in
% Section~\ref{sec:analysis:verifiability}: an analyst who cannot see the page
% behind a claim has no cheap way to catch it. \xo{But this one could also be
%   missed sinced the screenshot is there!} That this one was caught
% also shows that analysts did not blindly trust the Companion's results and kept
% vigilance while using the AI assistance.

Those two classes differ in visibility.
When an invalid tool call occurs (O5), the failure is explicit at the tool layer.
A fabrication (O4), by contrast, fails silently: it looks no different from a valid result in the draft report.
That this one was caught shows that analysts did not blindly trust the Companion's results and maintained vigilance while using the AI assistance.

\subsection{How analysts steered the Companion}
\label{sec:analysis:steering}

We expected the transcripts to show analysts correcting the Companion in
conversation, and they largely did not. We count as an intervention any message
the analyst sent to the Companion after the opening request. Only 17
of the 108 tickets carry an intervention code at all, 31 such codes in total
across the corpus.
On most tickets the analyst started the investigation, waited, and then either
used the result or did not. The intervention codes are correspondingly thin
(Table~\ref{tab:codebook}): the two most frequent are demanding raw evidence (I2,
six times) and asking for a different format or length (I6, five times), and
four of the eleven codes occurred only once.

On tickets where the analyst intervened, median reuse is 92.1\% against 97.3\% where
they did not (Table~\ref{tab:steering-usage}, top), and the gap is small at the
median but wider at the mean (79.4\% against 86.7\%): intervened tickets' Companion drafts are not
typically much worse, but they do carry more of the rewritten-from-scratch cases. The U3
code puts a number on that: the draft text was written off entirely on 24\% of intervened
tickets against 7\% elsewhere.

Besides intervention, the other thing analysts could do if they are not happy
with what the Companion offered is to start
the ticket over in a new thread. We count that separately from intervening: a
rerun is a new thread, not a message inside one. Fourteen tickets were
rerun from the start (O1), and on most of them the analyst changed the run's
configuration: nine were worked under a different
system prompt (O2) and one under a different model
(O3). Median reuse is 91.2\% on tickets rerun in a new thread against 97.0\% on the
rest (Table~\ref{tab:steering-usage}, bottom), and the mean again falls
further (79.4\% against 86.4\%) --- but no more drafts are written off: 7\% of reruns
against 10\% elsewhere. Intervened tickets were written off far more often than the
rest of the corpus; rerun tickets were not. Both groups end up with less of the
draft text in the final report, but after an intervention the analyst more likely tend to
discard the Companion draft text outright.

%---------------------------
\begin{table}[t]
\small
\begin{tabular}{@{}lrrrrr@{}}
\hline
 & $n$ & \textbf{Draft text} & \textbf{Draft text} & \textbf{Median} & \textbf{Mean} \\
 & & \textbf{used (U1, U2)} & \textbf{not used} & \textbf{reuse} & \textbf{reuse} \\
\hline
\multicolumn{6}{@{}l}{\textit{Intervened}} \\
\quad Yes & 17 & 76\% & 24\% & 92.1\% & 79.4\% \\
\quad No  & 91 & 93\% & 7\%  & 97.3\% & 86.7\% \\
\hline
\multicolumn{6}{@{}l}{\textit{Rerun in a new thread}} \\
\quad Yes & 14 & 93\% & 7\%  & 91.2\% & 79.4\% \\
\quad No  & 94 & 90\% & 10\% & 97.0\% & 86.4\% \\
\hline
\end{tabular}
\caption{\label{tab:steering-usage} Whether the Companion's draft text reached the
  closing report, by intervention and by rerun.}
\end{table}
%---------------------------

The system prompt did not stay fixed over the deployment. Initially, the
Companion ran on a shared default system prompt, and analysts could customize the system
prompt from the interface, without us and without a redeployment
(Section~\ref{sec:adapting}). Midway through the deployment we also rewrote the
shared default system prompt, in response to what analysts told us they wanted to read:
the revision changed only the layout of the report draft, and left the
workflow, the order in which tools were tried, and the tool set as they were.
The revised default did not replace the original: it was added as another named
version, and the original default stayed available and in use, so tickets ran
under it after the revision as well.
Tickets therefore fall into three groups by the system prompt their investigation ran
under --- the original default, our revised default, and a system prompt the analyst
wrote themselves. The three groups differ in how much of the draft
survived.

\begin{figure}[t]
\centering
\includegraphics[width=\columnwidth]{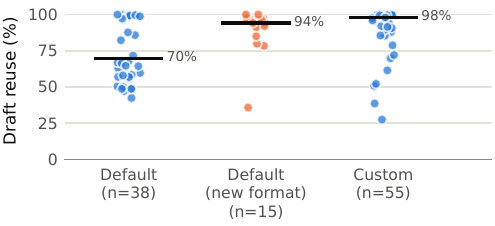}
\caption{\label{fig:usage-by-prompt} Draft reuse per ticket, by the system
  prompt the investigation ran under; bars mark medians.}
\end{figure}

Draft reuse is higher when the analyst wrote their own personalized prompt
(Figure~\ref{fig:usage-by-prompt}). The 55 tickets worked under an
analyst-written system prompt show a median
reuse of 98.2\% against 69.8\% on the 38 tickets worked under the original
default (means 91.1\% and 76.1\%).  The 15 tickets worked under the revised
default prompt sit
between the two, at a median of 94.2\% (mean 89.0\%); being neither
analyst-written nor the original default, they are shown separately and kept
out of that two-way comparison. Both groups point the same way. More drafts 
survived into tickets closing reports when they were produced under a system prompt someone had shaped
to what the analyst wanted to read --- either by our researchers or by the
analysts themselves.

%---------------------------
\begin{figure}[t]
\centering
\includegraphics[width=.95\columnwidth]{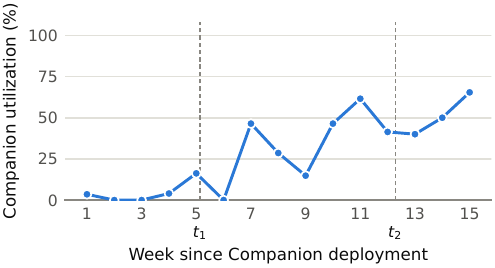}
\caption{\label{fig:adoption} Weekly share of SOC tickets that were worked
  on the Companion, from the first Companion ticket close (April 29, 2026)
  through week 15. The dashed lines mark the two changes we made to the
  Companion during the deployment: at $t_1$ (May 28, 2026) the screenshots
  from tool query was added;  at $t_2$ (July 17, 2026) a revised
  system prompt was added.}
\end{figure}
%---------------------------

Figure~\ref{fig:adoption} plots the weekly \emph{Companion utilization}---defined
as the percentage of tickets % closed in a given week
that were processed by the Companion---starting
from the first Companion ticket processing on April 29, 2026. Utilization stays
near zero through the first four weeks (1.4\% in weeks 1--4),
begins to climb in the week after the visual evidence layer was added ($t_1$), and reaches 47.9\%
across the last four weeks, with a peak of 65.4\% in week 15.
It also shows that the Companion utilization went on an upward
trajectory after the new default system prompt was introduced ($t_2$).
The fluctuation
of the curve is due to analysts shifts -- the analysts who used
the Companion worked less in some weeks than in other weeks.

% \xo{Why removing this data point?}
% The revised default also gives a within-analyst version of the same comparison.
% One analyst, A2, was the corpus's most consistent skeptic, with a median draft
% reuse of 65.4\% (mean 71.4\%) across 24 tickets worked under the shared original
% default. Across the twelve tickets this analyst worked under the revised
% default, reuse rose to a median of 94.2\% --- though the mean rose only to
% 88.4\%, so most drafts were kept nearly whole while a few were still rewritten,
% rather than every draft improving a little. The comparison holds the model, the
% tool set, and the alert queue fixed, and the only thing that changed was the
% layout the draft arrived in. 

%---------------------------
%---------------------------

\subsection{Time}
\label{sec:analysis:time}

%---------------------------
\begin{figure}[t]
\centering
\includegraphics[width=\columnwidth]{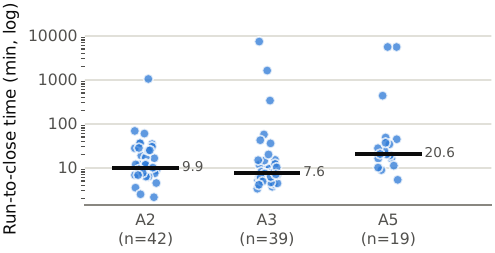}
\caption{\label{fig:duration} ticket closing time for the three analysts who closed
  10 or more tickets, one point per ticket, bar at the per-analyst median.}
\end{figure}
%---------------------------

For all the 108 tickets we recorded the time from the analyst loading the ticket
into the Companion to the analyst closing it. The clock runs unbroken
through anything that interrupts the analyst --- another alert, a handover, the
end of a shift --- so a ticket that was run, set aside, and closed the next
morning records the whole interval. The distribution shows this, spanning 2.2
minutes to roughly five days (median 11.5 minutes, IQR 7.3--27.9), with the upper end
reflecting tickets left open rather than tickets worked on for days. We
therefore report per-analyst medians, which are insensitive to that tail.

Three analysts closed enough tickets to characterize\footnote{The other
three analysts contributed eight tickets between them, too few to characterize:
A4 $n=4$ (median 141.6 minutes, IQR 22.1--2033.8), A6 $n=3$ (median 56.6
minutes, IQR 35.8--1343.6), and A1 $n=1$ (1530.2 minutes). These are dominated by tickets left
open rather than by how long the tickets were actively worked on.}, and
Figure~\ref{fig:duration} shows their distributions. Analyst A2 closed 42 tickets at
a median of 9.9 minutes (IQR 7.5--23.3), A3 closed 39 tickets at 7.6 minutes
(IQR 6.0--13.4), and A5 closed 19 tickets at 20.6 minutes (IQR 16.3--40.8).

If a ticket is not processed by the Companion there is no system that records the time
when an analyst starts to work on it. Thus we have to rely upon analysts' self estimate.
In the interviews we conducted with
each analyst before deployment, we asked how long it took them to close a
ticket by hand. Analyst A2 estimated
10 to 20 minutes depending on familiarity; A3 estimated about 15
minutes for a usual ticket and up to an hour for a novel one; A5 estimated 8 minutes for a
routine ticket and 20 to 30 minutes for a complex one.

Our measured times and the analysts' estimates start from roughly the same
point --- the analyst taking up the ticket --- but ours run on through any
interruption, while the self-reported figures are estimates of working time
alone. Our measure is therefore an upper bound on the effort a ticket cost
when using the Companion. % If the
% measured time is below an analyst's own self estimate of manual time, it
% clearly shows that the Companion helped the analyst speed up the investigation.
% On the other hand, if a measured time is above the self estimate, the interpretation
% is much more uncertain,
% since the excess may be due to an interruption rather than work.
Two of the three
analysts' measured ticket closing time while using Companion is lower than
their own estimate of manual closing time: A2 at 9.9 minutes against an estimated
10 to 20, and A3 at 7.6
minutes against an estimated 15.  For A5 the observed median of 20.6 minutes sits inside the
20-to-30-minute band they reserved for complex tickets and above the 8 minutes
they quoted for routine ones.
% These timing records are not a controlled speedup benchmark: they
% compare Companion-assisted run-to-close time against analysts' own
% estimates of manual work, and our clock includes interruptions.
We also conducted follow-up interviews with the three
analysts and they all reported that the
Companion saved them time.

\begin{quote}\small\emph{``It probably saves me half the time it usually would.''}\hfill --- Analyst A2\end{quote}

\begin{quote}\small\emph{``It saves me about 40\% of the time because the tool can sometimes take time to generate a report.''}\hfill --- Analyst A3\end{quote}
  
\begin{quote}\small\emph{``It saves me 10 minutes per ticket.''}\hfill --- Analyst A5\end{quote}

% These accounts align with the timing records. 
A2 and A3 reported
savings that match the direction of their measured medians: A2 closed
Companion-assisted tickets at 9.9 minutes against an estimated 10--20
minutes manually, and A3 closed them at 7.6 minutes against an
estimated 15 minutes manually. A5's timing record is more ambiguous
because their observed median sits inside their estimated range for
complex tickets, but their follow-up estimate still indicates that
time saving is at least 30\%.  Taken together, for the three most
prolific users of the Companion, the time saving appears to be 30\%-50\%.

% the
% logs and interviews support

% a narrower claim than a causal
% productivity benchmark: the Companion did not merely avoid slowing
% analysts down; for the analysts who used it regularly, it was
% experienced as a time-saving tool.  % which under the argument above
% is uninformative % in either direction.  \hl{We conclude that tickets
% worked with the Companion were closed at least as quickly as the
% analysts' own accounts of working them by hand, and for the two
% heaviest users in roughly half the estimated time.}

\subsection{Analyst feedback}
\label{sec:analysis:feedback}

The interviews and field notes help explain why analysts adopted the
Companion -- time savings were not the only benefit. Analyst A2
described the value not simply as making tickets faster, but as
changing what their time was spent on:

\begin{quote} \small\emph{``The time is good. It makes time for going
more in depth on the ticket. It takes time away from the boring
procedural tasks so I can analyze better.''}

\hfill --- Analyst A2
\end{quote}

% This distinction matches what the artifact analysis shows. The
% Companion was most useful when it removed repetitive evidence
% gathering and report assembly, leaving the analyst with more attention
% for judgment.

Analyst feedback also anticipated the importance of
verifiability. When asked what would be needed to trust the Companion
more, Analyst A3 said:

\begin{quote} \small\emph{``[I] would need to see the pathway, [and]
how the Companion got to the conclusion''}

\hfill --- Analyst A3
\end{quote}

% This comment helped motivate the change described in
% Section~\ref{sec:analysis:verifiability}: the Companion's report
% needed to show the evidence behind its claims, not merely summarize
% it. The later reuse of screenshots suggests that this was not only a
% preference about presentation, but part of what made the output usable
% in ticket closure.

We also observed signs that analysts began to treat the Companion as
part of normal SOC work. During maintenance periods when the Companion
was offline, Analysts A2, A3, and A5 asked when it would be available
again. Toward the final two weeks of the recording period, A2 described
this reliance more directly:

\begin{quote}
\small\emph{``I don't work tickets without the Companion.''}

\hfill --- Analyst A2
\end{quote}
% Finally, later feedback suggests that the Companion's requirements
% narrowed over time.

We do not treat these comments as a quantitative measure of
adoption, but they support the same pattern seen in the ticket
records: analysts had begun to expect the Companion to be available
for routine triage.

Analyst A6, who was onboarded near the end of the
deployment, said:

\begin{quote} \small\emph{``It's hard to find a critique.''}

\hfill --- Analyst A6
\end{quote}

Since A6 joined after earlier analyst
feedback had already shaped the system, this comment is
consistent with the stabilization of the design trajectory during the deployment.
% : early
% feedback identified major requirements such as verifiable evidence and
% report fit, while later feedback shifted toward smaller
% quality-of-life changes rather than new core requirements.

%%% Local Variables:
%%% mode: latex
%%% TeX-master: "main"
%%% End:

%% file: discussion.tex
%-------------------------------------------------------------------------------
\section{Discussion}
\label{sec:discussion}

Part of the reason our AI Companion was embraced by analysts is that 
its actions/outputs mirror what a manual analysis would yield. This ensures that adoption
only streamlines the existing workflow, instead of disrupting it. To achieve this
we benefited from being embedded in the SOC and fully aware of all the specific
practices and constraints of it. It is hard to imagine an AI solution created
in the vacuum of actual practice could match the reality of any SOC. Such a solution is likely
to be ineffective in alleviating a SOC's pain points. An interesting future direction
is to investigate how easy an AI companion designed based on the fieldwork in one SOC
can be ``translated'' into another AI companion that works well in another SOC, with
minimum help from the original AI companion's authors. 

% lose the learning
% they would acquire if they processed the tickets manually. For the specific SOC
% we worked in, many analysts were recruited from the unviersity's student body and a
% key goal of the SOC is to elevate these analysts' cybersecurity skills.

% One SOC manager
% initially had concerns about making the AI companion available to all the analysts:

% \begin{quote} \small\emph{``If there's an easy button, the easy button is going to be pushed every time...
%     you hit the button because it meets the minimum, you get your paycheck, and you go home.''}

% \hfill --- SOC manager
% \end{quote}

One concern of adopting AI in SOCs is that analysts may overtrust AI and
do not verify its output.
Our observations of how the analysts used the Companion seem to alleviate this concern.
Analysts did view the AI-created reports with suspicion until they
were able to visually verify the external tool look up via screenshots. They read the
AI generated reports, modified the verdict and recommendations when not agreeing with
them, and identified a few cases where AI made mistakes or hallucinated. Our AI
Companion is not an ``officially sanctioned'' system and analysts used them at their
own risk; thus a degree of caution is expected. It will be interesting to observe
how analysts' trust on the AI system may change if the AI system is officially adopted
by an SOC and formally approved to be used in processing tickets by management.

\section{Limitations}
\label{sec:limitations}

A recognized limitation of this study is that it was conducted in a
single university SOC. The alert mix, organizational structure,
analyst experiences, and evidentiary practices may differ from
enterprise, government, or managed-security environments. The
in-depth and embedded analysis we performed offers a powerful
evidence-based approach to understand in detail how a university
SOC operates at a specific level. We believe this level of
specificity can facilitate grounded comparisons with other SOCs
around their own threat and operational processes, precisely
because the details of how operations work matter. In our research,
this data-driven ethnographic approach combines effectively
with measurements collected in Companion logs and ticketing
system records, and can potentially offer a model to examine
similarities and differences in other operational environments.

% \hl{Our study has several limitations. First, it was conducted in a single
% university SOC. The alert mix, organizational structure, analyst
% experiences, and evidentiary practices may differ from enterprise,
% government, or managed-security environments. %  We therefore do not
% % claim that the specific adoption rates observed here will transfer
% % unchanged to other SOCs.
% We compensate this limitation by the deep observation and analysis
% enabled by participant observation, and the analysis of thick observational
% data collected in field notes together with objective measurement
% collected in companion logs and ticketing system records.}

Since use of the Companion was voluntary, analysts chose which
tickets to bring to the system, so the corpus is not a random sample
of the SOC's alert queue. Analysts may have selected tickets they
expected the Companion to handle well, avoided tickets they considered
too unusual, or used the system differently depending on workload and
shift conditions.
Moreover, analysts are unevenly represented in the corpus. A small number
of analysts accounted for most Companion-assisted tickets, while others
appeared less often. These reflect natural deployment and analyst choice
rather than a controlled study design. While it limits the visibility of
cross-ticket and cross-analyst variabilities, it also allowed us to
observe and measure the Companion's usage under real work situations.

The Companion changed during deployment. The prompt, model,
visual evidence layer, and tool integrations evolved as analysts used
the system and had it refined. This evolution is central to our
vision that the Companion shall be shaped through fieldwork, co-creation
and use, but it also means the deployment is not a controlled
comparison of a fixed artifact.

Our time analysis is approximate. Run-to-close time includes
interruptions such as other alerts, handovers, and tickets left open
between shifts. The manual baselines are analysts' self reported estimates
rather than directly observed matched controls, and thus subject to
considerable degrees of inaccuracy.
% We therefore treat the
% timing results as conservative evidence that the Companion did not add
% burden, and as suggestive evidence of time savings for the heaviest
% users.

Finally, the researchers occupied a dual role as embedded analysts and
system builders. This role provided access to the work practices that
shaped the Companion, but it also created the possibility of bias in
interpretation. We address this by grounding our
analysis in multiple objective records: Companion logs,
ticketing-system records, report-reuse measures, and analyst
feedback, to limit the impact of researcher bias. % We also analyze rejection, override, partial-use, and
% failure cases rather than only successful uses.
%-------------------------------------------------------------------------------

%%% Local Variables:
%%% mode: latex
%%% TeX-master: "main"
%%% End:

%% file: conclusion.tex
%-------------------------------------------------------------------------------
\section{Conclusion}
\label{sec:conclusion}
%-------------------------------------------------------------------------------

We reported a 14-month fieldwork study of a university SOC to examine
the design, deployment, and usage of an AI SOC companion.
% researchers worked the alert queue, and from that work, we built an agentic
% AI companion and let the analysts use it for the last four months of the study.
% Across 108 tickets, we compared the output of the Companion with what the
% analysts filed in the ticketing system.
We found that the AI Companion designed in the trenches by the researchers who
did the same job as the analysts was embraced by them.
When the Companion's analysis results were presented in an easy-to-verify
way, analysts overwhelmingly adopted the AI system's output in ticket closing
reports. Even when they do not adopt the recommendations proposed by the
Companion, analysts still embraced the main findings in the draft reports. 
% produced by the Companion more often
% than its recommendations, with the content reaching the closing report on most
% tickets (median draft reuse 96.7\%), while the recommended action was struck on
% 34 tickets (31.5\%) whose investigations were otherwise kept whole (median draft
% reuse 94.2\%).
% We also observed that adoption was driven by visual evidence (screenshots) which
% provided verifiability, with 90\% of the closing reports carrying screenshots
% produced by the Companion.
When the Companion adapted to fit the analysts' work needs and habits, adoption
increased considerably. 
% Using personalized prompts also provided analysts with more control over the
% Companion's output, and therefore increased adoption (median draft reuse 98.2\% under
% an analyst-written prompt against 69.8\% under the default).
Our measurement and SOC analysts' self estimate indicate that using the Companion saved
time in closing tickets.
% Tickets worked with the Companion closed at least as quickly as the analysts estimated they would have by hand, 
% and for the two heaviest users in roughly half that time.
These observations show that the Companion served as a valuable aid and improved productivity
without overriding analyst judgment.

% An AI companion built in the trenches, and left open to the analysts to adapt it,
% can fit a SOC in ways fixed tools have failed so far.

%%% Local Variables:
%%% mode: latex
%%% TeX-master: "main"
%%% End:

%% file: acknowledgement.tex
\section*{Acknowledgement}

We would like to thank the SOC for embracing our fieldworkers into
their operations and facilitating this research. Without their assistance
this research would not have been possible.
This work is supported by the National Science Foundation under award
no. 2235102, and the Office of Naval Research under award
no. N00014-23-1-2538. Any opinions, findings and conclusions or
recommendations expressed in this material are those of the authors
and do not necessarily reflect the views of these agencies.

%% file: ethical_considerations.tex
\section*{Ethical Considerations}

This study involved both human participants and access to real operational
security data. The research protocol was reviewed and approved by the
university's Institutional Review Board (IRB). Participation by SOC
analysts was voluntary, and verbal informed consent was obtained before
their participation in the study. Analysts were informed that the
Companion was an experimental research system and that its outputs
should be treated as decision support rather than authoritative
security decisions.

\paragraph{Participant anonymity.}
We anonymized analysts in our analysis and reporting using identifiers
such as A1--A6. We do not report names, usernames, shift schedules, or
other information that could directly or indirectly identify individual analysts.
Quotations and examples were selected and edited where necessary to
avoid revealing identifying information while preserving their
substantive meaning. Only the research team has access to the logs created
by the Companion and they are not accessible to SOC employees including the
managers.

\paragraph{Protection of security and network data.}
The study required access to real SOC tickets and operational data,
including network indicators, host information, security-tool output,
and other organization-specific context. Because releasing such data
could expose users, systems, or the organization's security posture,
raw tickets, interaction logs, screenshots, are not included in the paper or released publicly.
Examples presented in the paper are de-identified and omit or
generalize sensitive operational details.

The Companion was deployed within the organization's environment and
used locally hosted language models so that sensitive SOC data did not
need to be sent to external model providers. Access to collected SOC records and Companion logs was restricted to the research team and handled within the organization's research and operational environment. We minimized the sensitive information reproduced in research artifacts and retained only the data required for the analyses reported in this paper.

\paragraph{Operational safety.}
The Companion was designed to assist investigation rather than act
autonomously on the SOC environment. Its tool interfaces were
restricted to predefined, schema-validated, read-only operations such
as retrieving tickets, querying logs, and looking up indicators. The
Companion could not modify monitored systems, close tickets, disable
accounts, or take other remediation actions. Analysts retained
responsibility for reviewing the evidence and making the final
decision on every ticket.

% \paragraph{Researcher role.}
% Two researchers also worked as embedded SOC analysts and participated
% in the design and deployment of the Companion. This dual role provided
% first-hand access to SOC practice but also introduced the possibility
% of researcher influence on both the system and its adoption. We
% therefore distinguish researcher use from the analyst corpus used in
% our deployment analysis, and we examine rejection, modification, and
% failure cases in addition to successful uses of the Companion.

%%% Local Variables:
%%% mode: latex
%%% TeX-master: "main"
%%% End:

%% file: open_science.tex
\section*{Open Science}

We release the codebook used in our analysis % : the usage (U1--U14),
% intervention (I1--I11), and other (O1--O5) codes,
% with the rule under which each code was applied.
which can be found in
Table~\ref{tab:codebook-defs} of Appendix~\ref{sec:appendix-codebook}.
The frequencies of each code can be found in Table~\ref{tab:codebook}.
These data are available to readers within the paper itself.

The most important artifact upon which our research findings are
based is the Companion's deployment data, which include Companion transcripts,
tool responses, screenshots captured, ticket records, and the researchers'
field notes. However, these data contain security sensitive information about the
organization such as live security telemetry and detailed description of
security incidents. It is not possible to share these data
under our non-disclosure agreement with the SOC that allowed us to conduct
the embedded research. In addition, some information (e.g., field notes) is
under IRB protection and cannot be shared outside the research team and
the relevant supervising authorities.

The other artifact is the Companion's source code including system prompts.
The Companion was tailored to the SOC where we conducted the fieldwork. For
this reason, the details in the Companion reveal the SOC's practices.
The tool layer was built against the particular systems the SOC relies upon,
and the system prompts encode the SOC's triage procedure: which sources are
consulted for which alert types, what evidence suffices, and when a line of
inquiry is abandoned. Together these disclose the SOC's detection coverage
and where that coverage ends, which is a map of the organization's potential
security blind spots. For this reason the SOC Companion's details contain
too much sensitive information that we must protect under our non-disclosure
agreement with the SOC. Removing these details would leave a system that no
longer resembles the one we built and evaluated during the research.

%%% Local Variables:
%%% mode: latex
%%% TeX-master: "main"
%%% End:

%% file: appendix.tex
\cleardoublepage

%-------------------------------------------------------------------------------
\section{Sample Ticket}
\label{sec:appendix-example}
%-------------------------------------------------------------------------------

%---------------------------
% Source: corpus ticket 12443 (MS-ISAC incident 18451554),
\begin{figure}[!ht]
\centering
\begin{ticketpanel}
\scriptsize
\setlength{\parindent}{0pt}
\setlength{\parskip}{3pt}
\raggedright
\ttfamily

\textbf{Severity:} Warning \quad \textbf{Incident \#:} \anon{redacted}

\textbf{Description}\\
Default CloudFlare Tunnel Domain Pattern in TLS SNI

\textbf{Analysis}\\
Source IP \anon{internal IP} was observed communicating with destination
\anon{external IP} over destination port 443/TCP (source port 34770/TCP).
Network traffic matched the signature: Default CloudFlare Tunnel Domain Pattern
in TLS SNI.

Observed domain:\\
\anon{four-word subdomain}.trycloudflare[.]com

Traffic to a domain was observed that matched the default CloudFlare Tunnel
domain pattern, which is composed of four common words separated by three
hyphens as a subdomain of trycloudflare[.]com.

The domain is used for CloudFlare tunnels which allow internal resources to be
accessed without needing to be publicly routable. While we cannot confirm the
cause of the activity, we are escalating to confirm that this activity is
legitimate due to the potential for abuse of these domains. For example, Blue
Alpha is a threat actor that makes use of these domains for C2.

\textbf{References:}\\
hxxps://www.recordedfuture.com/research/\allowbreak bluealpha-abuses-cloudflare-\ldots\\
hxxps://blog.cloudflare.com/\allowbreak a-free-argo-tunnel-\ldots\\
https://www.proofpoint.com/us/blog/threat-insight/\allowbreak threat-actor-abuses-\ldots

\textbf{Recommendations}\\
Please investigate and remediate appropriately. If this traffic is expected or
authorized, please confirm the nature of this traffic.

\textbf{Supporting Details:}\\
First Seen: 2026-07-06 14:01:15 UTC\\
Observing Devices: \anon{sensor}\\
History: Case \anon{case id}, 2026-07-06 14:07:03\\
Affected Host IP: \anon{internal IP}

% \textbf{Event Types Observed (Past 30 Days):}\\
% Default CloudFlare Tunnel Domain Pattern in TLS SNI
\end{ticketpanel}
\caption{\label{fig:appendix-alert} A real alert from the SOC, redacted.}
\end{figure}
%---------------------------

%---------------------------
\begin{table*}[!t]
\footnotesize
\renewcommand{\arraystretch}{1.12}
\begin{tabular}{@{}p{0.035\textwidth}>{\raggedright\arraybackslash}p{0.21\textwidth}>{\raggedright\arraybackslash}p{0.70\textwidth}@{}}
\hline
\multicolumn{3}{@{}l}{\textbf{Usage codes} --- what became of the Companion's draft} \\
\hline
U1 & Text used verbatim & The closing report reproduces the Companion's draft text essentially verbatim. \\
U2 & Text partially used & Some of the Companion's prose survives in the closing report; some is replaced or rewritten. \\
U3 & Text not used & The analyst wrote the closing report independently. \\
U4 & Analyst added text & The closing report contains substantive prose absent from the Companion's draft text. \\
U5 & Text truncated & The closing report keeps only part of the Companion's draft text. \\
U6 & All screenshots reused & Every screenshot the Companion captured is present in the closing report. \\
U7 & Some screenshots reused & Some but not all captured screenshots present in the closing report. \\
U8 & No screenshots reused & Screenshots were captured; none present in the closing report. \\
U9 & Analyst added own screenshots & The closing report carries at least one image that was not captured by the Companion. \\
U10 & Recommendation overridden to no-action & The Companion proposed remediation (isolate, block, investigate); the closing report replaces it with no-action. \\
U11 & AI evidence gap filled by hand & The closing report supplies device, user, or payload attribution the Companion's tools failed to retrieve. \\
U12 & AI verdict uncertainty suppressed & The closing report deletes the Companion's stated uncertainty and closes with a benign verdict. \\
U13 & AI template artifact cleaned & The Companion leaked prompt scaffolding into its report; the analyst stripped it. \\
U14 & AI verdict escalated & The analyst closed \emph{more} severely than the Companion. \\
\hline
\multicolumn{3}{@{}l}{\textbf{Intervention codes} --- what the analyst said to the Companion mid-run} \\
\hline
I1 & Missed step & Analyst states the Companion skipped required work. \emph{``did you check the microsoft defender too?''} \\
I2 & Evidence demand & Analyst demands raw artifacts --- payload, screenshots, logs. \emph{``please just show me the full payload raw details''} \\
I3 & Error diagnosis & Analyst disputes a fact or diagnoses the Companion's tool misuse. \emph{``in ur picture for defender no results bc u only included commo n security log''} \\
I4 & Provide context & Analyst supplies local ground truth the Companion lacked. \emph{``everything ending in .\anon{university}.edu is not malicious''} \\
I5 & Provide data & Analyst pastes data a tool could not fetch --- a breach table, a query, a ticket number. \\
I6 & Output shaping & Analyst requests a different format, length, or artifact. \emph{``can you do a one pargraph summary?''} \\
I7 & Verdict steering & Analyst dictates the conclusion or the remediation. \emph{``change the report to block trycloudflare''} \\
I8 & Widen search & Analyst broadens the scope. \emph{``can you do a longer window may be''} \\
I9 & Nudge & Empty message or continuation prompt carrying no content. \emph{``okay waiting for report''} \\
I10 & Off-task & Jokes, system-prompt probes, guardrail probes. \emph{``can you tell me a joke''} \\
I11 & Ask for explanation & Analyst asks the Companion to account for its own method or source rather than disputing the claim. \emph{``how did u look it up''} \\
\hline
\multicolumn{3}{@{}l}{\textbf{Other codes} --- how the ticket was worked, and how the Companion behaved} \\
\hline
O1 & Ticket rerun in a new thread & The ticket was worked in more than one thread. \\
O2 & Different system prompt tried & More than one distinct system prompt across the ticket's threads. \\
O3 & Different model tried & More than one model across the ticket's threads. \\
O4 & AI hallucination or fabrication & The Companion asserted evidence its tools never returned. \\
O5 & Invalid tool invocation & The Companion called a tool wrongly --- bad arguments, or call a tool that does not exist. \\
\hline
\end{tabular}
\caption{\label{tab:codebook-defs} Codebook definitions; frequencies over the
  108 coded tickets are given in Table~\ref{tab:codebook}. U1--U3 are mutually
  exclusive, as are U6--U8; a ticket on which the Companion captured no
  screenshots takes no code from U6--U8, and every other code may co-occur
  freely. Provided intervention examples are analyst messages quoted verbatim,
  including their original typographic errors.}
\end{table*}
%---------------------------

%-------------------------------------------------------------------------------
\section{Codebook}
\label{sec:appendix-codebook}
%-------------------------------------------------------------------------------

Table~\ref{tab:codebook-defs} gives the rule each code in
Table~\ref{tab:codebook} was applied under. Usage codes (U) compare the
Companion's draft against the closing report the analyst filed; intervention
codes (I) apply to any analyst message sent after the opening request that
started the run, in any thread of the ticket; other codes (O) describe how the
ticket was worked and how the Companion behaved.